\documentclass[sigplan,nonacm]{acmart}
\usepackage[english]{babel}
\usepackage{amsmath}
\usepackage{mathtools}
\usepackage[noend]{algpseudocode}
\usepackage{ifthen}
\usepackage{textcase}
\usepackage{framed, newfloat}
\usepackage{xspace}
\usepackage{comment}
\usepackage[normalem]{ulem}
\usepackage{enumitem}
\setitemize{noitemsep,topsep=0pt,parsep=0pt,partopsep=0pt}
\setlist{noitemsep,topsep=0pt,parsep=0pt,partopsep=0pt}
\setlist[itemize]{leftmargin=*}
\usepackage{listings}
\usepackage{color}
\usepackage{multirow}
\usepackage[linesnumbered, ruled, vlined]{algorithm2e}
\usepackage[font=footnotesize, textfont=normalfont]{caption}

\graphicspath{{figures/}}

\definecolor{dkgreen}{rgb}{0,0.6,0}
\definecolor{gray}{rgb}{0.5,0.5,0.5}
\definecolor{mauve}{rgb}{0.58,0,0.82}
\usepackage{cleveref}
\crefname{section}{}{\S\S}
\crefdefaultlabelformat{\S#2#1#3}

\newcommand{\sys}{Xumi\xspace} 
\newcommand{\comprehend}{\textit{Intent Comprehension}\xspace} 
\newcommand{\conflict}{\textit{Conflict Detection \& Resolution}\xspace} 
\newcommand{\deploy}{\textit{Deployment Optimization}\xspace} 
\newcommand{\campus}{\textit{CampusNet}\xspace}
\newcommand{\cloud}{\textit{CloudNet}\xspace}
\newcommand{\extreme}{\textit{ExtremeNet}\xspace}

\renewcommand\footnotetextcopyrightpermission[1]{} 
\setcopyright{none}

\begin{document}
\title{Automating Conflict-Aware ACL Configurations \\ with Natural Language Intents}

\author{%
Wenlong Ding\textsuperscript{1},
Jianqiang Li\textsuperscript{1},
Zhixiong Niu\textsuperscript{2},
Huangxun Chen\textsuperscript{3},
Yongqiang Xiong\textsuperscript{2},
Hong Xu\textsuperscript{1}
}

\affiliation{%
\vspace{4pt}
\textsuperscript{1}The Chinese University of Hong Kong, \textsuperscript{2}Microsoft Research, \\
\textsuperscript{3}Hong Kong University of Science and Technology (Guangzhou)
\country{}
}

\renewcommand{\shorttitle}{}
\renewcommand{\shortauthors}{}

\begin{abstract}
ACL configuration is essential for managing network flow reachability, yet its complexity grows significantly with topologies and pre-existing rules. 
To carry out ACL configuration, the operator needs to (1) understand the new configuration policies or intents and translate them into concrete ACL rules, (2) check and resolve any conflicts between the new and existing rules, and (3) deploy them across the network.
Existing systems rely heavily on manual efforts for these tasks, especially for the first two, which are tedious, error-prone, and impractical to scale.

We propose \sys to tackle this problem. 
Leveraging LLMs with domain knowledge of the target network, \sys automatically and accurately translates the natural language intents into complete ACL rules to reduce operators' manual efforts.
\sys then detects all potential conflicts between new and existing rules and generates resolved intents for deployment with operators' guidance, and finally identifies the best deployment plan that minimizes the rule additions while satisfying all intents. 
Evaluation shows that \sys accelerates the entire configuration pipeline by over 10x compared to current practices, addresses $O$(100) conflicting ACLs and reduces rule additions by  $\sim$40\% in modern cloud network.
\end{abstract}

\maketitle

\section{Introduction} 
\label{sec:intro}

An access control list (ACL) is an ordered list of rules deployed at a router interface to control access to network resources~\cite{acl-cisco,acl-h3c,acl-huawei}. It has long been used as a basic security mechanism in network management~\cite{jinjing,netconfeval}. 
An ACL rule has five attributes: source and destination IP prefixes, protocol/port, time range (optional), and action (permit or deny), that specify precisely what to do for packets matched to this set of conditions.
An ACL is processed in a top down manner, i.e. the action of the first matched rule is applied to a packet and the processing ends.

ACL configuration is a foundational task in network management.
Currently it heavily relies on manual effort~\cite{lumi,jinjing}. 
Network operations teams (NetOps) start by reasoning about a new policy or intent expressed in natural language. 
For example, for university NetOps, a new intent may be, ``We shall block access to ChatGPT websites during the final exam period in all exam areas to prevent cheating''.
NetOps must reason about the intent and translate it into valid ACL rules with attribute specifics like OpenAI's IP prefixes.  
Then they determine the deployment plan of which ACLs to add these new rules to, assess the impact on existing rules, and verify if the rules and deployment plan are correct.
Finally they configure routers using vendor-specific scripts to deploy the new rules.
Except for the final step where automation tools can assist~\cite{robotron2016,openconfig-ref,apstra-gdb,YANG}, most of this process is manual. 

Not only is this process tedious and error-prone, but it also has become a bottleneck for the evolution of network management at scale. 
Alibaba reports that deploying one new intent in their cloud WAN can take about a week~\cite{jinjing}. 
Our interview with another cloud WAN NetOps reveals similar configuration times. 
They also share that new intents can be triggered by daily configuration changes, and a large volume of them may arrive simultaneously when new devices are added.

Automating ACL configuration with natural language intents poses three key challenges. 
\begin{itemize}
    \item \textit{Reasoning from Intent to Rule.} 
    Large language models (LLMs), with their unprecedented language capabilities, seem a perfect fit for translating natural language intents into ACL rules to reduce manual effort. 
    This allows NetOps to focus solely on reviewing and approving the rules generated by LLMs. Yet, LLMs have inherent limitations when used directly here. They lack proprietary network-specific information for attribute reasoning, such as translating ``exam areas'' into the correct IP prefixes. 
    They also suffer from hallucinations that produce incorrect outputs. These limitations can lead to additional manual effort and corrections which negate the automation benefits.

    \item \textit{Detecting Conflicts with Existing Rules.}  
    New intents may conflict with some existing rules if they need to be deployed on the same ACL, which must be resolved before deployment. This can happen a lot in a large network with many existing ACL rules. 
    Manual conflict detection is often impractical due to the combinatorial complexity, even for smaller campus networks. As a somewhat alarming example, we learned from a university NetOp that they have to resort to a reactive deploy-then-fix method to only resolve conflicts when they are reported by users, underscoring the need for proactive and automated conflict detection to prevent service loss before deployment.

    \item \textit{Optimizing the Deployment.} 
    NetOps eventually deploy new ACL rules to fulfill intents with as few rule additions as possible to ensure ease of maintenance. A common strawman approach places rules on all interfaces along the paths of the flows pertaining to an intent. 
    This creates many redundant rules: for instance, a ``deny'' intent can be enforced by adding the rule to just one interface along a path. 
    Due to the interaction between different rules of the same ACL and multiple ACLs on the same path, how to globally minimize total rule additions while ensuring correctness is a difficult combinatorial problem for multiple intents, each with multiple rules.
\end{itemize}

Recent work has started to look into similar problems~\cite{lumi,jinjing,netconfeval}. 
Lumi~\cite{lumi} considers intent-based network configuration, but is limited to extracting natural language fragments such as ``exam areas,'' leaving the core reasoning task, such as mapping exam areas to the corresponding prefixes or gateway interfaces, to NetOps. It also ignores the subsequent conflict detection and deployment optimization tasks.
JINJING~\cite{jinjing} considers actual deployment but relies on low-level inputs like endpoint prefixes and router interfaces which are tedious and absent in high-level intents. 
Moreover, it only focuses on intent achievements without considering potential conflicts between new and existing rules. 
NetConfEval~\cite{netconfeval} uses LLMs to directly generate configurations without specific designs for hallucination mitigation or conflict detection. It relies completely on iterative human feedback to mitigate LLM's incorrect outputs, which remains labor-intensive and hard to scale to large networks.

In this work, we build \sys, a novel automated system for ACL configuration  with minimal human intervention. \sys tackles the three challenges with the following design. 

First, \sys performs {\comprehend} using LLMs to help NetOps translate natural language intents into valid ACL rules for subsequent deployment (\cref{moti:intent} and \cref{design:comprehend}). 
We aim to maximize correctness of LLM's output in order to reduce manual feedback on correcting errors like prior works~\cite{lumi,netconfeval}.
Specifically, we maintain a Semantics-Network Mapping Table (SNMT) to supply LLMs with up-to-date network-specific information required by ACL, and employ various prompting techniques to mitigate potential hallucinations. Once LLM's outputs, expressed rigorously in an intermediate representation, are reviewed and approved by NetOps, the generated new rules are ready for subsequent tasks.

Next, \sys proactively conducts {\conflict} to address all potential conflicts between each new intent rule and all existing rules. 
We show that the detection problem is more involved than it seems, due to many complicating factors such as the impact of preceding ACL rules and routing paths that can result in false positives (\cref{moti:detect}). 
Thus we first propose a novel concept named truly matched flows, representing flows exactly matched by a rule and excluding those matched by its predecessors, to ensure accurate individual detection for each new and existing rules. We further validate if a conflict flow routes on a feasible path to ensure detection accuracy.
Note to resolve a conflict, \sys as a general system needs to provide maximum flexibility to NetOps instead of constraining them into a fixed set of strategies. Thus \sys takes a conservative approach that allows NetOps to propose a resolution intent in natural language for each intent, if necessary, which are then also translated automatically to produce the reconciled rules (\cref{design:conflict}).

Finally, \sys runs {\deploy} to identify a plan with minimal rule additions to implement all intents.
Starting from current practices~\cite{jinjing,acl-cisco} such as \textit{endpoint} deployment (\cref{moti:deployment}), we first present a few non-trivial scenarios where rule additions can be further reduced without affecting correctness, including using one rule to cover not only its corresponding intent but potentially also some others (\cref{moti:deployment}). 
We formulate an optimization program that incorporates these possibilities to address the combinatorial problem across all new intents (\cref{design:deploy}).

We prototype \sys on various LLMs and evaluate it in networks ranging from small campus (41 routers) to modern cloud (171 routers) to extreme-scale network $\sim$6x larger than a typical cloud (\cref{eval}). 
Results show that \sys with GPT-4o achieves 98.5\% and 90\% comprehension accuracy without NetOps feedback for campus and modern cloud networks respectively, and fully comprehends all intents within 8 minutes with at most 3 feedback iterations. 
\sys detects conflicts with 3.33x higher accuracy than baseline methods, completing detection in at most 2 hours for extreme-scale networks. Additionally, it reduces deployed rules by 38.8\% on average compared to baselines, with deployment decisions made in under 3 minutes. 

\section{Motivation and Challenges} 
\label{sec:moti}

We start by outlining the novel research problems in \sys's three tasks, and our key insights on how to address them.

\subsection{Intent Comprehension}
\label{moti:intent}

\sys first interprets the new natural language intents and derives the corresponding ACL rules with valid attributes~\cite{acl-cisco,acl-h3c,acl-huawei} and gateway information as accurately as possible, thereby minimizing NetOps' manual involvement. 
For the running example of blocking ChatGPT in exam areas during the final period introduced in~\cref{sec:intro}, the five attributes include the \textit{source} (prefixes for OpenAI's ChatGPT service), \textit{destination} (say ``10.0.1.0/24'' and ``10.0.2.0/24'' for the school's ``exam areas''), \textit{application} (TCP ports 80 and 443 for web), \textit{time range} (weekdays in May 2025), and \textit{action} (deny). 
Enumerating all unique combinations of source, destination, and application then generates the corresponding ACL rules for this intent. 

Naturally, we leverage large language models (LLMs) \cite{GPT,touvron2023llama,jiang2023mistral} for intent comprehension.
LLMs have shown superior language processing capabilities compared to their predecessors~\cite{ner1,ner2,ner3,vague1,vague2} like BERT~\cite{bert}, and their large-scale pretraining on massive corpora also teaches them general computing knowledge including network protocols. 
They also easily handle the diverse natural language expressions for the same attribute as shown in Table~\ref{tab:intent_comprehend} without NetOps handcrafting rigid pattern matching rules. 
These advantages make LLMs an ideal candidate with the potential for high or even perfect comprehension accuracy. However, LLMs do suffer from two intrinsic limitations that hinder their accuracy when directly applied to our task.

\begin{table}[t]
     \centering
     \large
     \renewcommand{\arraystretch}{1.5}
     \resizebox{1.0\columnwidth}{!}{
\begin{tabular}{|c||c|c|}
\hline
Attributes & ACL Config Specifics & NL Fragment \\ \hline
Source & 104.18.33.170/32, 184.105.99.79/32, ... & OpenAI, oai, openai, ... \\ \hline
Destination & 10.0.1.0/24 & Laboratory, lab, labs, ... \\ \hline
Protocol(Port) & TCP(80), TCP(443) & HTTP and HTTPS, websites, web service ... \\ \hline
Time Range & weekdays & Mon. to Fri., business days, ... \\ \hline
Action & deny & prohibit, forbid, disallow, ban, ... \\ \hline
\end{tabular}
}
    \vspace{5pt}
     \caption{Examples of \comprehend. 
     ACL configuration has rigid specifications for each attribute. 
     An attribute specific may be expressed in multiple ways in natural language.}
     \vspace{-18pt}
     \label{tab:intent_comprehend}
\end{table}

$\bullet$~\textbf{Network-Specific Information.} 
Understanding ACL or network configuration intents requires not only interpreting the text but also incorporating the domain-specific information from the underlying network. 
For example, an LLM can easily identify ``OpenAI'' as the source from an intent as in Table~\ref{tab:intent_comprehend}, but it lacks domain-specific knowledge about (1) the prefixes of OpenAI's web services, and (2) the router interfaces in the given network that can reach OpenAI. Unlike general knowledge (e.g., HTTP/HTTPS over TCP) learned during pre-training, this network-specific information is proprietary and keeps changing over time, and has to be explicitly incorporated. 

$\bullet$~\textbf{LLM Hallucinations.} 
Hallucination is a well-known open issue with LLMs and currently an active research topic in the AI community~\cite{hallucination1,hallucination2,LLMConfig,netconfeval}.
Compared to more tolerant tasks like email drafting, intent comprehension in network configuration has stringent correctness requirements on LLM's output. 
Thus, we need to prevent LLMs from, for instance, fabricating a list of some TCP port numbers when the intent involves \textit{any} TCP traffic, or misidentifying TCP as the protocol when the intent is for all traffic (due to the context where all prior intents target TCP, leading the LLM to believe that ``TCP'' is just omitted here), which we have actually observed in our experiments.

\noindent\textbf{Main Ideas.} To address these two key issues and achieve high comprehension accuracy, we first introduce a \textit{Semantics-Network Mapping Table (SNMT)} that provides up-to-date network-specific information required for creating valid ACL rules in the target network (e.g., entity prefixes and gateway interfaces) (\cref{design:comprehend:ir_snmt}). Next, we design an \textit{IR-enhanced System Prompt} to rigorously format ACL attributes into a unified Intermediate Representation (IR) (\cref{design:comprehend:ir_snmt}) and use prompting techniques such as \textit{Chain-of-Thought Reasoning}, \textit{Few-Shot Demonstration}, and \textit{Self-Reflection} to guide IR completion and validation against hallucinations (\cref{design:comprehend:hallucination}). 
For each LLM round, \sys invites NetOps to review and approve the output IR or provide feedback, which is organized into a \textit{Feedback Prompt} for iterative refinement until final approval is obtained (\cref{design:comprehend:feedback}). \sys finally generates valid ACL rules from the approved IR for subsequent tasks (\cref{design:comprehend:generation}).

\subsection{Conflict Detection \& Resolution} 
\label{moti:detect}

\begin{figure}[t]
	\centering
	\includegraphics[width=1.0\columnwidth]{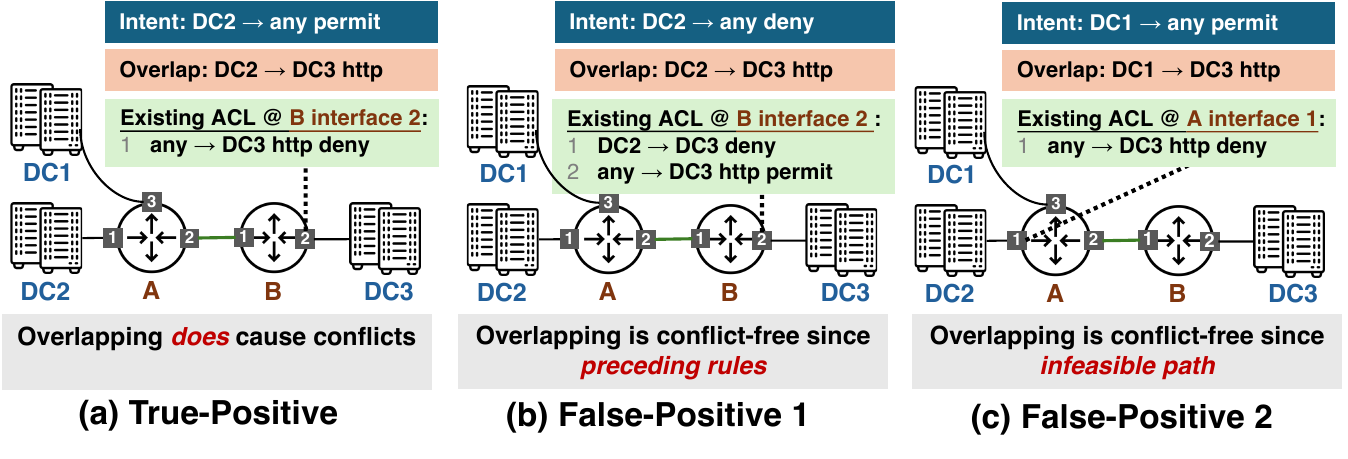}
 \vspace{-18pt}
	\caption{Conflict detection. Naive method detecting flow overlaps between intents and existing rules with opposite actions can cause false positives.}
\vspace{-10pt}
 \label{fig:conflict_situations}
\end{figure}

After \comprehend, each intent can be characterized by a \textit{group} of rule(s) that may need to be added to different ACLs. 
A new rule, however, may conflict with existing rules in the same ACL, requiring NetOps to explicitly resolve them before deployment to ensure correct actions are enforced on the flows as mentioned in~\Cref{sec:intro}.
We do assume, without loss of generality, that the new intents themselves in the same round of configuration are conflict-free (as a result of thorough negotiation among the stakeholders). 

At first thought, conflict detection simply involves checking two conditions between a new rule and existing ones in the target ACL: (1) whether there is an overlap between source-destination pair, application, and time range; and (2) whether they apply opposite actions on the overlapping segments. Figure~\ref{fig:conflict_situations}(a) provides an example: the new rule permits traffic from DC2 to any destination, but along the routing path from DC2 to DC3, i.e., DC2$\rightarrow$A$\rightarrow$B$\rightarrow$DC3, an existing rule on router B (interface 2) denies any HTTP traffic to DC3. This creates a conflict for HTTP traffic from DC2 to DC3.

What makes conflict detection intriguing is that, these two conditions alone may lead to many false positives (FPs), and neglecting them results in an $\sim$70\% accuracy drop as shown in our evaluation (\cref{eval:detect}).

$\bullet$~\textbf{Case 1: FPs due to Preceding Rules.} 
In Figure~\ref{fig:conflict_situations}(b), when examining the new rule ``DC2$\rightarrow$any deny'' for potential conflict with the ACL on router B's interface 1, there appears to be a conflict with the second existing rule ``any$\rightarrow$DC3 http permit'' for DC2 to DC3 HTTP traffic. 
Yet, since ACL's processing order is from top to bottom, the first rule ``DC2$\rightarrow$DC3 deny'' takes priority and would have already applied the same deny action as the new rule over DC2$\rightarrow$DC3 traffic, which means the previously detected conflict using the two conditions is a FP.

$\bullet$~\textbf{Case 2: FPs due to Infeasible Routing Paths.} 
Another case of FPs arises when the interface with conflicting existing rules is not on a viable routing path for the detected overlapping traffic. 
As shown in Figure~\ref{fig:conflict_situations}(c), although the new rule ``DC1$\rightarrow$any permit'' and the existing rule at router A's interface 1's ACL ``any$\rightarrow$DC3 http deny'' have opposite actions for DC1 to DC3 HTTP traffic, router A's interface 1 is not on any path from DC1 to DC3. 
Therefore, this detected conflict has no actual effect and is a FP. 

\noindent\textbf{Main Ideas.} \sys tackles FPs in conflict detection rigorously. 
First, to avoid FPs caused by preceding rules, we define \textit{truly-matched flows} to precisely characterize traffic matched to a rule at its position \textit{after} all preceding rules fail in ACL. 
Conflicts are then detected by overlapping these truly-matched flows with the intent traffic. 
Second, \sys uses \textit{interface-path validation} to eliminate FPs caused by infeasible paths, by verifying whether routing paths of conflict flows include the conflicting interface (\cref{design:conflict:accurate}). 
Finally, to resolve conflicts after detection, \sys invites NetOps for proposals to \textit{protect} flows that should retain their existing rules, called protect intents.
\sys then automatically generates resolved intent rules for conflict-free deployment (\cref{design:conflict:resolution}).
We cannot resort to a fixed set of criterion to resolve conflicts, since when to follow the existing rules and when to follow the new ones depend on many idiosyncrasies such as business considerations. 

One might wonder if network verification tools~\cite{batfish,jinjing,Minesweeper,NetDice,QARC,ARC,ProbNetKAT} such as Batfish~\cite{batfish} could be used for conflict detection, as they validate flow reachability by directly identifying interfaces where reachability is violated. 
While verification tools can detect whether flows specified by an intent have the opposite intent actions, their focus on binary reachability results limits their effectiveness. It is sufficient for them to identify the first path and interface where a flow's reachability is violated, whereas conflict detection requires identifying \textit{all} potential violations simultaneously to ensure resolution. Without a full view of all violations, resolving one conflict may easily introduce new ones, creating a long or even infinite resolution loop. Furthermore, intents often specify a large number of flows, and applying this process to each flow also significantly increases the complexity, rendering verification-based methods impractical.

\subsection{Deployment Optimization} 
\label{moti:deployment}

\begin{figure}[t]
	\centering
	\includegraphics[width=1.0\columnwidth]{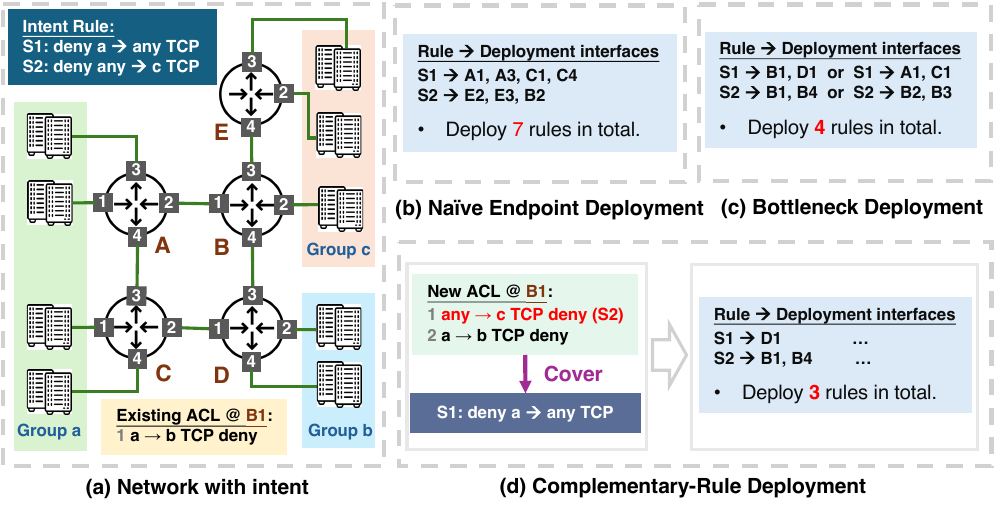}
 \vspace{-17pt}
	\caption{Deployment optimization methods of intent rules.}
 \label{fig:deny_depl}
 \vspace{-8pt}
\end{figure}

The final step is to deploy conflict-free rules by inserting them at the top of target ACLs, following the default and common practice~\cite{acl-cisco,jinjing}. 
The deployment requirements clearly differ for ``permit'' and ``deny'' rules: a ``deny'' rule is satisfied as long as it is applied to at least one interface along each possible routing path, while a ``permit'' rule must be deployed on all interfaces of all possible paths. 
Thus, \sys deploys ``deny'' rules at the \textit{path level} to at least one ACL per path, and ``permit'' rules at the \textit{interface level} on all ACLs along the traffic's paths. 
Specifically, we take an \textit{endpoint deployment} method that places ``deny'' rules on either source or destination gateway of each path, depending on which has the least total number of interfaces. We interviewed NetOps from a university and a production, and they confirmed these as the most privileged methods for ACL rule deployment.

While these simple approaches ensure correctness, they often introduce redundant rules, increasing the complexity of subsequent updates and maintenance, which NetOps are striving to avoid. As our evaluations show (\cref{eval:deploy}), they result in 36.3\% and 41.4\% redundant rule additions for ``permit'' and ``deny'' intents respectively, compared to \sys's designs leveraging novel deployment observations below across all evaluated cases. 

$\bullet$~\textbf{Bottleneck Deployment for ``Deny'' Rules.} 
We first note that endpoint deployment for ``deny'' rules does not always minimize rule additions, as it overlooks \textit{path-level} flexibility of placing the rule on any interface along a path. 
For example, in Figure~\ref{fig:deny_depl}(b), {endpoint deployment} adds 7 rules by placing rule S1 on all source interfaces and S2 on all destination interfaces for two intents. 
In contrast, deploying rules at ``bottleneck'' interfaces, where most paths traverse through, reduces rule addition to 4 for the same effect as in Figure~\ref{fig:deny_depl}(c). 
Placing S1 on interfaces B1 and D1 (or A1 and C1) satisfies the requirement that each path has at least one rule, equivalent to deploying it on all source interfaces.

$\bullet$~\textbf{Complementary-Rule Deployment.} 
Our second observation is, surprisingly, adding a new rule may even fulfill another intent in some cases, creating more opportunities for us to reduce the rule additions needed. 
For example, at interface B1 in Figure~\ref{fig:deny_depl}(d), the new rule S2 ``any$\rightarrow$c tcp deny'' combined with the existing rule ``a$\rightarrow$b tcp deny'' covers the traffic managed by S1 ``a$\rightarrow$any tcp deny''. 
Deploying S2 on interface B1 thus eliminates the need for S1 on the same interface, further reducing the total rule additions to 3 from 4 based on bottleneck deployment in Figure~\ref{fig:deny_depl}(c). Obviously, this principle also applies to ``permit'' rules if the actions of intent and existing rules are reversed.

\noindent\textbf{Main Ideas.}  
We strive to minimize total rule additions for a given set of intents by leveraging the two observations above, 
which interact globally across interfaces and intents. 
For a ``deny'' intent $I$ for instance, the interface covering the most paths for $I$ may not be the one that maximizes the coverage of other intents deployable at the same interface with its existing rules.
We thus formulate optimization problems to capture the combinatorial nature and determine the best deployment plan (\cref{design:delpoy:formulation}). 
To do this we rely on a novel concept of \textit{Equivalent Intent Set}: for each intent $I$, on each interface $J$ it traverses, we identify all possible intents each of which can cover $I$ with complementary rules, and group them into an {equivalent intent set} for that intent $I$ and interface $J$ (\cref{design:deploy:eq}).

Recently JINJING~\cite{jinjing} also considers reducing rule additions in ACL updates, highlighting its importance in production networks. 
However, it focuses on minimizing new rules on a single interface after the rules to be deployed are predetermined. 
For example, if ``any$\rightarrow$a deny'', ``b$\rightarrow$any deny'', and ``deny all'' are to be deployed on one interface, JINJING suggests only deploying ``deny all'' as it covers the first two. This simple optimization aligns with our second observation. 
\sys further considers the global impact of complementary rules in determining which interfaces to be used for rule deployment, further minimizing potential rule additions. 

\section{Design Overview} 
\label{sec:workflow}

 \begin{figure}[t]
	\centering
	\includegraphics[width=1.0\columnwidth]{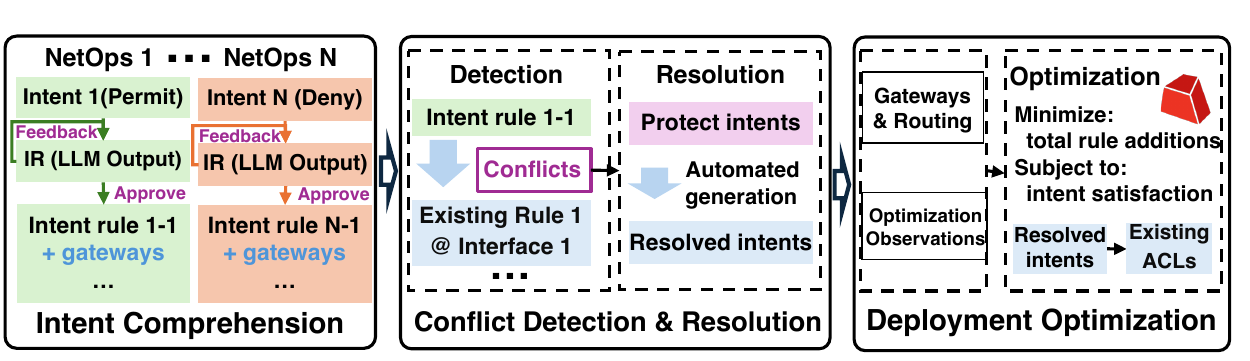}
    \vspace{-20pt}
	\caption{\sys architecture.}
 \label{fig:overview}
 \vspace{-10pt}
\end{figure}

\sys automates conflict-aware and deployment-efficient ACL configuration from natural language intents. Figure~\ref{fig:overview} illustrates its architecture with three sequential modules.
\begin{itemize}
    \item First, \comprehend module (\cref{design:comprehend}) processes NetOps' natural language intents and translates them into valid ACL rules with source and destination router gateways. NetOps are always involved in reviewing and approving LLM outputs (IR in~\cref{moti:intent}) in this process. If errors are found during review, NetOps provide feedback for LLMs to refine the outputs until correct. \sys assumes that NetOps provide conflict-free intents and considers potential conflicts between newly generated rules and existing ones. 
    \item Second, \conflict module (\cref{design:conflict}) proactively identifies these potential conflicts across all interfaces along the routing paths of traffic specified by each new rule, and returns the specific set of conflict flows. NetOps are then invited to resolve them by identifying which (conflict) flows shall be ``protected'' to retain actions in their existing rules, and \sys subsequently generates conflict-free intent rules for deployment. 
    \item Finally, \deploy module (\cref{design:deploy}) deploys all the resolved intent rules by inserting them into target ACLs. Leveraging source and destination gateways identified before with routing information in the network topology that we assume \sys knows, \sys determines the precise paths corresponding to each rule. It then leverages insights including bottleneck deployment and complementary rules to minimize the total number of rule additions while ensuring intent satisfaction, by formulating these considerations as an optimization problem.
\end{itemize}

\section{Intent Comprehension} \label{design:comprehend}

\begin{figure}[t]
	\centering
	\includegraphics[width=1.0\columnwidth]{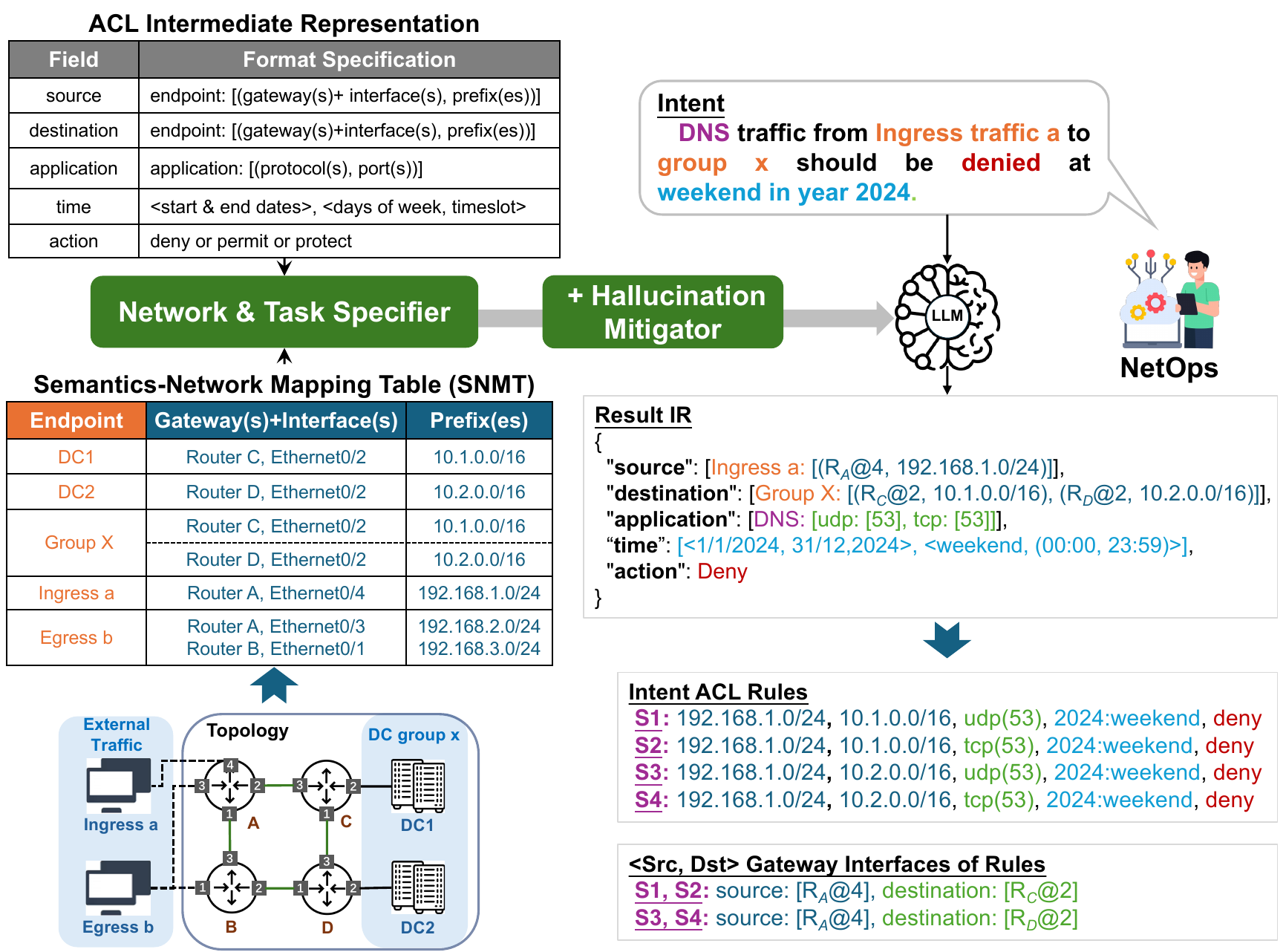}
	\caption{Workflow of \comprehend with an intent example.} 
 \label{fig:understanding_flow}
 \vspace{-5pt}
\end{figure}

Our goal is to instruct LLMs to understand NetOps' natural language intents and derive ACL rules in quintuple format with their endpoint gateways with high accuracy, minimizing human involvement. Figure~\ref{fig:understanding_flow} presents a design overview with a specific example, highlighting two key components.

$\bullet$~\textbf{Network \& Task Specifier} (\cref{design:comprehend:ir_snmt}): it incorporates carefully designed data structures to embed network-specific information unknown to LLMs but essential for ACL configuration, along with task-specific details to guide the LLMs in identifying the correct ACL attributes.

$\bullet$~\textbf{Hallucination Mitigator} (\cref{design:comprehend:hallucination}): it provides additional guidance to ensure LLMs follow the correct comprehension logic and revises its output if necessary. 

Lastly, we detail \sys's manual feedback mechanism (\cref{design:comprehend:feedback}) and methods to generate valid ACL rules with their gateways from LLM's output approved by NetOps (\cref{design:comprehend:generation}). 

\subsection{Network \& Task Specifier} \label{design:comprehend:ir_snmt}

As shown in Figure~\ref{fig:understanding_flow}, we design two key data structures: the \textit{Semantics-Network Mapping Table} (SNMT) for network specification, and the \textit{ACL Intermediate Representation} (IR) for task specification. These two components are parts of the \textit{system prompt} for LLMs.

\noindent\textbf{SNMT for Network Specification.} According to ACL domain knowledge~\cite{acl-cisco,acl-huawei,acl-h3c}, LLMs require all five ACL attributes for comprehension but lack real-time topology details, which are absent from both LLMs' knowledge and NetOps' intents (recall \cref{moti:intent}). In Figure~\ref{fig:understanding_flow}, LLMs can infer attributes like \textit{application} (e.g., DNS with port 53 for TCP and UDP) but lack knowledge of prefixes and gateways (e.g., for ``group x'') in the target network. To address this, \sys maintains a Semantics-Network Mapping Table (SNMT), which maps natural language entities to: (1) gateway routers and interfaces for determining routing paths for rule deployment, and (2) IP prefixes for ACL rule generation. For example, ``DC1'' maps to gateway ``R$_C$@2'' (Router C, interface 2) and prefix ``10.1.0.0/16''. Note that LLMs can handle diverse entity expressions (e.g., ``DC1'', ``Datacenter 1'' or even typos like ``datecetre 1''), so SNMT does not need exhaustive entries for these variations. SNMT by default pairs a gateway interface with the corresponding prefix, but also allows enumeration between them when required, such as ``egress b'' which is mapped into a single pair of two gateways and prefixes, as both prefixes can egress through R$_A$ and R$_B$.

In practice, real-time running metadata for devices or device groups, such as their names, IPs, gateways, group IDs, and other specifics, is maintained in a dataset, as confirmed through our interviews with NetOps in production. 
So we use automated scripts to extract the information required by SNMT to ensure its timeliness and avoid the need for tedious and frequent manual updates by NetOps.

\sys uses SNMT as an external information source with prompting~\cite{prompt1,prompt2} rather than fine-tuning~\cite{NetLLM}. 
We make this decision due to SNMT's dynamic nature, which requires frequent updates in practice to reflect current network state. 
This makes fine-tuning resource-intensive and susceptible to outdated results caused by LLMs' hallucination. When SNMT changes, \sys can flexibly reset the context and reload the \textit{system prompt} with the updated SNMT for comprehension.

\noindent\textbf{ACL IR for Task Specification.} 
\sys then organizes extracted ACL attributes into an IR with template depicted in Figure~\ref{fig:understanding_flow} for subsequent rule generation. 
The \textit{source} and \textit{destination} fields record all gateway-prefix pairs for an endpoint as a list. 
Natural language name of the endpoint is also included for NetOps review. 
Similarly, \textit{application} field captures application name with its protocol-port pairs as a list. 
The \textit{action} field includes ``permit'', ``deny'', or ``protect'' which we introduce in~\cref{design:conflict:resolution}. 
If NetOps do not explicitly specify an element in IR, say time, \sys puts ``any'' as the default.

\subsection{Hallucination Mitigator} 
\label{design:comprehend:hallucination}

Besides SNMT and IR, we use additional techniques below in \textit{system prompt} to overcome LLM's hallucinations.

\noindent\textbf{Chain-of-Thought Reasoning.} 
Inspired by~\cite{chainofthought}, we provide step-by-step instructions to guide LLMs in locating the extraction source of each IR element based on the intent, SNMT or LLM's general knwoledge. 
Specifically, endpoint names, time ranges, and ACL actions are derived from NetOps' inputs, with application protocols and ports inferred from LLMs' general knowledge and network-specific details from SNMT. 
We utilize five-step instructions for filling each of the five IR fields; details are given in \Cref{app:prompt_example} Figure~\ref{fig:prompt_skeleton}(a) for brevity. 

\noindent\textbf{Few-Shot Demonstration.} 
We further include a few examples of intents with their ground-truth IR to emphasize corner cases that LLMs are prone to error. For example, while LLMs typically associate a single protocol and port within an IR, DNS can use both UDP and TCP; this dual-protocol case is such an example.

\noindent\textbf{Self-Reflection.} 
We also design verification steps asking the LLM to verify identified IR elements. 
Specifically, the LLM checks for typical hallucination cases: (1) \textit{Non-existent elements}, where IR elements fall outside valid range given in SNMT or absent from the intent, prompting LLMs to revisit related reasoning steps such as identifying source/destination endpoints or application fields; 
and (2) \textit{Incorrect element dependency}, such as assigning a wrong port to a protocol that does not support it, requiring LLMs to redo the application protocol inference step. 
Figure~\ref{fig:prompt_skeleton}(a) in \Cref{app:prompt_example} illustrates how these verification steps integrate into the \textit{chain-of-thought reasoning} prompts to enhance IR correctness. 

\subsection{NetOps Feedback} \label{design:comprehend:feedback}
\sys always invites NetOps to review the LLM's output after each inference round, either approving it for subsequent tasks or providing feedback on errors. The feedback is expected to specify what LLM outputs incorrectly and how to find correct answer. This feedback, combined with wrong output and initial intent, is organized into a \textit{feedback prompt} and returned to LLM for iterative refinement until result is correct and approved. Detailed prompt skeleton and NetOps-LLM interaction examples are provided in~\Cref{app:prompt_example}.

\subsection{Intent Rule Generation From IR} 
\label{design:comprehend:generation}

After obtaining the IR from the LLM's output for a given intent, the corresponding ACL rules can be generated from the IR. 
Since each ACL rule requires a single source and destination prefix with a single protocol (with or without a port), \textit{source}, \textit{destination}, and \textit{application} fields are enumerated to produce all possible rules. For the running example in Figure~\ref{fig:understanding_flow}, two destination prefixes and two protocol-port pairs yield four new rules.

\sys also saves the gateway and interface information for source and destination of each rule, to facilitate subsequent rule deployment. 
This can be readily obtained by looking up the SNMT using the IP prefixes. 

\section{Conflict Detection \& Resolution} \label{design:conflict}

Our next step is to identify and resolve conflicts between each new rule and all existing ones. 
The discussion here begins with accurate detection by avoiding the two false positive cases introduced in~\Cref{moti:detect} (\Cref{design:conflict:accurate}). 
We then explain how to resolve conflicts with NetOps' ``protect'' intents (\Cref{design:conflict:resolution}).

\subsection{Accurate Conflict Detection} \label{design:conflict:accurate} 

We first propose the concept of \textit{Truly-Matched Flows (TMF)} of each existing rule to ensure detection is unaffected by the impact of preceding rules. 
We then conduct \textit{Interface-Path Validation} to confirm that detected conflicting flows traverse through the feasible paths and constitute a real conflict. 

\noindent\textbf{TMF-Based Conflict Detection.} 
This process has two phases: \textit{TMF Calculation} and \textit{Intent Overlapping}. 
In the first phase, TMF of a given rule at an ACL's position $k$ (from the top), denoted as $T_k$, is defined as the subset of its specified flows (without considering other rules) $R_k$ excluding those that are already matched by the preceding rules. 
Formally, we calculate from the top of an ACL using $T_k = R_k - \bigcup_{i=1}^{k-1} T_i$, initialized with $T_1 = R_1$\footnote{Each interface's ACL includes a default ``permit'' or ``deny'' rule for unspecified flows by other rules, which \sys also accounts for during detection.}. Recall example in Figure~\ref{fig:conflict_situations}(b) in~\Cref{moti:detect}: TMF of the 2$^\text{nd}$ rule includes all HTTP flows to DC3, except those originating from DC2. Then for a new intent rule $I$, conflicts are identified by checking the overlap between $I$'s traffic $R_I$ and $T_k$ of each existing rule in the target ACL with opposite actions. 
This approach avoids FP case 1 in~\Cref{moti:detect} and is applied to ACLs across all interfaces traversed by $I$'s traffic. 

We comment that implementation of the above detection process is non-trivial due to the \textit{set operations} on flows such as \textit{difference}~($-$) for {TMF calculation} and \textit{intersection}~($\cap$) for {intent overlapping}. 
Each rule may use prefixes at different granularity, making it hard to efficiently represent the results of set operations.
We refer the reader to \Cref{sec:implement} for more details, but the key idea is that, for each rule, we expand its flow set to the most fine-grained level found in the SNMT: e.g. when the rule contains DC1's prefix, we expand its flow set into all clusters of DC1 as specified in SNMT for conflict detection. 

\SetAlCapFnt{\small}
\SetAlCapNameFnt{\small}
\begin{algorithm}[t]
\caption{\small Interface-Path Validation of Conflict Flows}
\label{algo:ipv}
\footnotesize
\KwIn{$All\_C$: Conflict flows detected at all interfaces; $SNMT$.}
\KwOut{$All\_VC$: Validated conflict flows on all interfaces.}
\SetKwFunction{FMain}{Validate\_Each\_Flow}
\SetKwProg{Fn}{Func}{:}{}
\Fn{\FMain{$interface, flow, SNMT$}}{ \label{line:ipvstart}
    \tcp{Obtain all routing paths with gateways in SNMT}
    $i_{src},i_{dst}=$ Match$(flow.src, flow.dst, SNMT.all\_prefix)$\; \label{line:entry}
    $paths =$ Routing$(SNMT[i_{src}].GWs$, $SNMT[i_{dst}].GWs)$\; \label{line:route}
    \If{interface $\in paths$}{ \label{line:match1}
        \Return True \tcp*{Interface-path validation}
    }
    \Return False \label{line:macth2}
}
\tcp{Validate conflicts on all rules of all interfaces}
$All\_VC = []$\; \label{line:validate1}
\For{(interface, rule, conflict\_flow) $\in All\_C$}{
    \If{\texttt{Validate\_Each\_Flow}(interface, conflict\_flow, SNMT)}{
        $All\_VC.\text{add}((interface, rule, conflict\_flow))$\;
    }
}
\Return $All\_VC$\; 
\label{line:validate2}
\end{algorithm}

\noindent\textbf{Interface-Path Validation.}  
For each conflict flow detected on an ACL, we validate whether it routes through this interface (lines~\ref{line:ipvstart}--\ref{line:macth2} in Algorithm~\ref{algo:ipv}) to address FP case 2 in~\Cref{moti:detect}. 
The flow's source and destination prefixes are matched against SNMT to identify gateway-prefix pair entries (line~\ref{line:entry}), where a match occurs if the prefix is equal to or a subset of the one in SNMT. Using these entries, we determine gateway interfaces of endpoints and retrieve all their feasible routing paths already known by \sys (line~\ref{line:route}). 
Finally, we check if the interface is used on any path to classify if it is a FP or not (lines~\ref{line:match1}--\ref{line:macth2}). This process is repeated for all conflict flows across existing rules on all interfaces, retaining only non-FP conflict flows as the final detection result (lines~\ref{line:validate1}--\ref{line:validate2}).

\subsection{Conflict Resolution} \label{design:conflict:resolution}

\sys presents all detected conflict flows to NetOps for explicit resolution. 
We introduce a simple ``protect'' mechanism, allowing NetOps to specify in natural language, which (conflict) flows should retain actions in their existing rules as another intent. 
These ``protect'' intents are comprehended in the same way as in~\cref{design:comprehend}, except the \textit{action} is now ``protect''. Note that \sys enables NetOps to express a broad ``protect'' intent covering multiple flows, instead of specifying for each individual conflict, to reduce manual effort. For example, rather than addressing conflicts like ``DC1$\rightarrow$DC2 TCP'' and ``DC3$\rightarrow$any TCP'' one by one, NetOps can define a general intent to ``protect'' all HTTP traffic. \sys then automatically overlaps protect flow set with actual conflicts to determine the ``protect'' flows for subsequent processing, such as ``DC1$\rightarrow$DC2 HTTP'' and ``DC3$\rightarrow$any HTTP''.

\sys then generates protect rules for these flows for deployment use, assigning them with anti-intent actions (e.g., ``deny'' for a ``permit'' intent) to preserve existing behaviors.
Each intent rule is then combined with protect rules proposed by the same NetOps to form a resolved intent, with protect rules placed above intent rule to correctly take effect.

\section{Deployment Optimization} \label{design:deploy} 

We now need to deploy the new rules to satisfy all resolved intents. 
Each resolved intent with its rules as in~\cref{design:conflict:resolution} is treated as a unit for deployment and we insert the rules at the top of the ACLs of the chosen interfaces. 
To leverage the two findings in~\cref{moti:deployment} for reducing the rule additions, we first compute the \textit{Equivalent Intent Set} (EIS) for each intent $i$, identifying whether another intent $j$ can satisfy $i$  when $j$'s rules are already deployed on the same ACL (\cref{design:deploy:eq}). 
We then formulate an optimization program to fully explore the effect of these equivalences in reducing rule additions across multiple intents, while balancing the trade-off of deploying ``deny'' rules at bottleneck interfaces (\cref{design:delpoy:formulation}). 

Given the different deployment natures of ``permit'' and ``deny'' intents at \textit{interface level} and \textit{path level} respectively as mentioned in~\cref{moti:deployment}, we group resolved intents by their original action for separate optimization. 
Our design below applies to either intent group independently.

\subsection{Equivalent Intent Set} \label{design:deploy:eq} 
To compute EIS for each intent, we first identify its equivalence computation scope by determining relevant interfaces and the associated intents that may be deployed on them.

\noindent\textbf{Equivalence Computation Scope.} 
For a resolved intent $RI_i$ in the group $\mathbb{RI}$ with the same original action, we identify its EIS on all interfaces where $RI_i$ has the potential to be deployed. 
Such an interface $p$ must satisfy two conditions: (1) flows of $RI_i$ that follow its original action, denoted as $RI_i^{\text{flow}}$, must traverse $p$, and (2) existing rules on $p$ should not include $RI_i$, necessitating a new deployment. 
For the first condition, we focus on $RI_i^{\text{flow}}$ as new rules only need to target flows and interfaces required for the original intent action, excluding those governed by ``protect'' rules in $RI_i$. For the second condition, since each interface's ACL enforces a default rule controlling all network flows (\cref{design:conflict:accurate}), essentially it is satisfied when $RI_i^{\text{flow}}$ shows conflicting actions between the intent and existing rules. This occurs when a conflict with $RI_i$'s original intent is detected at $p$, which can be determined from conflict detection results.

We use the binary variable $Q_{pi}$ to indicate if the two conditions above are met simultaneously. Only $RI_i$ and $p$ with $Q_{pi} = 1$ are considered for EIS calculation and subsequent decision formulations.
EIS computation of $RI_i$ is then limited to the set of interfaces $\{ p \mid Q_{pi} =1, \forall i\}$.

\noindent\textbf{Equivalence Condition.} To determine if a resolved intent $RI_j $ is equivalent to $RI_i$ on a given interface $p$, we leverage the complementary-rule observation in~\cref{moti:deployment} by checking whether the flow set $RI_j^{\text{flow}}$ together with the TMF of all existing rules with the same action on $p$ can cover $RI_i^{\text{flow}}$. 
Intents satisfying this condition are collected into $RI_i$'s EIS. 
Formally, let $T_{pk}$ represent the TMF of the $k^\text{th}$ rule on the ACL at $p$, $RI_i$'s EIS on $p$ can be computed as:
\[
E_{pi} = \{ RI_j \mid RI_i^{\text{flow}} \subseteq \cup_{k}^{T_{pk}.\text{act} = RI_j.\text{act}} T_{pk} \cup RI_j^{\text{flow}}, \forall RI_j \in \mathbb{RI}_p\}.
\]
Here, $\mathbb{RI}_p$ denotes the set of resolved intents with potential to be deployed at interface $p$, i.e., $\{RI_j \mid Q_{pj} = 1\}$. 
Note that $RI_i$ is always included in its own EIS $E_{pi}$, as it is equivalent to itself by definition. We have an accelerated implementation for this equivalence calculation (\cref{sec:implement}).

\subsection{Optimization Formulations} \label{design:delpoy:formulation}

We now present the optimization formulations based on EIS to minimize the total rule additions across resolved intents and network interfaces. 
In addition to EIS, the formulations take as input the rule count for each resolved intent $RI_i$ donated as $t_i$, the potential deployment indicator $Q_{pi}$ mentioned before, and path set $\Omega_i$ for $RI_i^{\text{flow}}$.
A deployment plan is specified by binary variables $N_{pi}$ for all $Q_{pi} = 1$, indicating whether $RI_i$ is deployed on $p$ or not. 

\noindent\textbf{Optimization for ``Permit''.} Essentially, the optimization task for the ``Permit'' group on interface $p$ ($\mathbb{G}_p$) is to find a sub-group of resolved intents with minimal number of rules that satisfies all intents in $\mathbb{G}_p$. 
As shown in Constraint~\eqref{eq2}, each intent $RI_i \in \mathbb{G}_p$ is satisfied by deploying at least one of its equivalent rules (including itself), i.e., $RI_j \in E_{pi}$, which can be achieved via a logical OR operation. 
With Objective~\eqref{eq1}, which sums the rule numbers for all deployed intents, the optimization program is able to explore all deployment combinations among intents and find one optimal solution. 
Formally, the formulation is defined as:
\begin{align}
    \min & \quad \sum_{i=1}^{|\mathbb{G}_p|} t_i N_{pi} \label{eq1} \\
    \text{s.t.} & \quad \bigvee_{RI_{j} \in E_{pi}} N_{pj} = 1, \quad \forall i, RI_i \in \mathbb{G}_p. \label{eq2}
\end{align} 
Since optimizations across interfaces are independent, the optimal decisions for each interface collectively form a globally optimal deployment plan.

\noindent\textbf{Optimization for ``Deny''.} We begin by considering the bottleneck interfaces alone first, as each path requires at least one deny rule to be installed (recall~\cref{moti:deployment}). 
We formulate this {path-level} constraint for each resolved deny intent $RI_i$ on each of its paths $P \in \Omega_i$ as $\bigvee_{p\in P} N_{pi} = 1$. 
Note that we should not consider path $P$ for $RI_i$ if $RI_i$ is already covered in at least one of $P$'s interfaces by the existing rules since the path-level nature of deny intents, i.e. only consider $P$ such that $\bigwedge_{p \in P} Q_{pi} = 1$.

Next, we incorporate EIS by revising the path-level constraint above: instead of deploying $RI_i$ directly, we allow deploying any equivalent rule $RI_j \in E_{pi}$ from its EIS. Specifically, we replace $N_{pi}$ in the path constraint $\bigvee_{p\in P} N_{pi} = 1$ with $\bigvee_{RI_j \in E_{pi}} N_{pj}$, enabling the optimization program to account for interactions between equivalence rules, as shown in Constraint~\eqref{eq4}. 
The complete optimization formulation for the group of ``deny'' intents $\mathbb{D}$ is presented below:
\begin{align}
    \min & \quad \sum_{i=1}^{|\mathbb{D}|}\sum_p  t_i N_{pi} \label{eq3} \\
    \text{s.t.} & \quad \bigvee_{p \in P} \bigvee_{RI_j \in E_{pi}} N_{pj} = 1, \label{eq4} \\
    & \quad \forall i\in \mathbb{D}, P \in \Omega_i \, \text{such that} \bigwedge_{p \in P} Q_{pi} = 1. \label{eq5}
\end{align} 

\section{Implementation} \label{sec:implement}

This section addresses key implementation issues of \sys.
The prompts used are anonymously open-sourced at~\cite{xumi}, and the full code will be released upon paper acceptance. 

\noindent\textbf{LLM Context Length Limitation.} 
Our first problem is for large networks, its SNMT can exceed  maximum context length of the LLM, preventing it from being processed in one prompt. 
To address this, we partition SNMT into slices and process them iteratively. 
Each slice, combined with \textit{system prompt} and IR from the last iteration, is fed into the LLM to generate an updated IR. Entities not identified in the current slice are labeled as ``Not Found'' in IR, and subsequent slices are processed until all endpoints in intent are found.

\noindent\textbf{Efficient Flow Set Operations.} 
Expanding each rule into flows based on the most fine-grained level of endpoint prefixes in SNMT (recall~\Cref{design:conflict:accurate}) can result in over millions of flows as observed in our evaluation (\Cref{eval}). 
Operating such large flow sets with attribute specifics (e.g., prefixes) is obviously computationally expensive and memory-intensive. 
To address this, \sys stores unique fine-grained flows with their attributes in a \textit{global flow information table}, and flow sets only record \textit{indices} referencing this table, reducing redundant storage and avoiding expensive set operations on attributes. 
These indexed flow sets are further encoded as \textit{Bitarrays}~\cite{bitarray} in Python with each bit representing the presence of a flow, enabling fast bitwise operations for set operations with further reduced storage and computation overhead. 
We also implement an efficient method to expand a rule to its indexed flow set, avoiding enumeration of all flows with attribute specifics and searching through the huge \textit{global flow information table}.
\Cref{app:bitarray} has more details. 

Furthermore, we employ an \textit{early termination} mechanism for flow set overlapping between intents and existing rules (\textit{Intent Overlapping} in~\Cref{design:conflict:accurate}). 
As soon as the accumulated TMF of all the preceding rules already cover the intent traffic, the current and subsequent rules can be skipped right away. 

\noindent\textbf{Accelerated EIS Calculation.} 
To check equivalence between (resolved) intents, \sys requires a \textit{union} of TMF for all existing rules with the same intent action on an interface (recall~\cref{design:deploy:eq}). Recomputing this for every intent pair is inefficient. 
Therefore, \sys precomputes aggregated flow sets for both ``permit'' and ``deny'' actions per interface upon finishing TMF calculation for all existing rules. 
These precomputed sets are reused during equivalence checks, eliminating redundancy and improving efficiency.

\noindent\textbf{Redundant Protect Rules Elimination.} 
While redundant intent rules can be eliminated through \textit{complementary-rule} observation (recall~\cref{moti:deployment}), overlapping intent rules may still have the same protect rules as they protect the same subset flows. When deployed in the same interface’s ACLs, this can also introduce redundancy. To address this, we remove duplicate protect rules in the updated ACL, retaining only the one with the highest priority.

\section{Evaluation} \label{eval}

In this section, we first present \sys's end-to-end performance and then evaluate the designs of its three modules in detail across various intents, network scales, and configurations. Our results reveal that: 

\begin{itemize}
     \item Overall, \sys completes end-to-end configuration of 20 intents within 20 minutes for a modern-scale data center (DC) network with 171 routers, addressing conflicts on 183 ACLs and reducing total rule additions by approximately 40\% compared to current practices (\Cref{eval:end_to_end}).
    \item \sys achieves 98.5\% and 90\% accuracy (with GPT-4o) in intent comprehension for campus and cloud networks, respectively, without manual feedback. All intents are correctly processed within 8 minutes with feedback (\cref{eval:comprehend}). 
    \item \sys delivers 3.3x higher accuracy than baseline methods ignoring any FP for conflict detection. For our largest network with over 10k interfaces and 2k rules per interface, detection can be done within 2 hours (\Cref{eval:detect}).
    \item \sys solves the deployment optimization within 3 minutes for the largest network, and reduces rule additions by >60\% in the best case and 38.8\% on average (\Cref{eval:deploy}).
\end{itemize}

\subsection{Evaluation Setup}
\label{subsec:eval_setup}
\begin{table}[t]
     \centering
     \renewcommand{\arraystretch}{1.35}
     \resizebox{1.0\columnwidth}{!}{

\begin{tabular}{|c||cccc||c|}
\hline
\multirow{2}{*}{} & \multicolumn{4}{c||}{Network Information} & \multirow{2}{*}{\begin{tabular}[c]{@{}c@{}}\#Prompt Tokens\\ (SNMT ratio)\end{tabular}} \\ \cline{2-5}
                        & \#Routers & \#Links & \#Interfaces & \#Entities &               \\ \hline
\campus  & 41        & 66      & 361          & 60         & 4407 (51\%)   \\
\cloud   & 171       & 219     & 2524         & 352        & 38755 (94\%)  \\
\extreme & 1026      & 1236    & 11796        & 2062       & 178414 (99\%) \\ \hline
\end{tabular}

}
    \vspace{5pt}
     \caption{
     Evaluated networks. We show token numbers of \textit{system prompt} and its SNMT proportions. For DC networks (\cloud and \extreme), entity numbers are shown at the cloud WAN level (each DC as a unit).
     }
     \vspace{-10pt}
     \label{tab:network}
\end{table}

We use OpenAI's GPT-4o (version 2024-05-13) and o1 (version 2024-12-17) via APIs from Azure OpenAI Service~\cite{GPT}. In addition, 
we locally deploy two open-weight LLMs, Qwen2.5-72B with Int4 quantization~\cite{Qwen} and Llama3.1-70B with FP8 quantization~\cite{llama}, provided by Hugging Face~\cite{huggingface}, running on four Ampere A100 GPUs with 80GB memory each in tensor parallel mode. 
\sys's remaining two modules are implemented in Python and run on three servers with four Intel Xeon Platinum 8268 24-Core processors and 768GB RAM each.
Gurobi Optimizer~\cite{gurobi} is used to solve the optimization. 

\noindent\textbf{Networks and Entities.}  
We evaluate \sys on three synthetic networks of varying scales as shown in Table~\ref{tab:network}, to reflect realistic topologies of a university and Azure's global DC network publicly available in~\cite{azure}. 
\campus's SNMT has entities of 9 departments, associated facilities such as office buildings, and grouped entities spanning multiple buildings and networks such as exam areas. 
\cloud models Azure's global DC topology~\cite{azure} comprising 57 regions with an average of 5 DCs per region, and includes 100 server groups (each with 16 servers) for two specific DCs to include the internal DC networks (DCN). The DCN topology is a 3-tier fat-tree~\cite{fattree1, fattree2}. 
\extreme scales \cloud by a factor of 6, replacing each region with 6 similar ones to represent networks of future scales. 
Each of these networks includes 10 external entities, each with 5 ingress/egress points at random gateway routers. 
Routing paths between each two routers are configured using k-shortest paths with k=4, following typical WAN settings~\cite{zhao2020smore,b4,swan}.

\noindent\textbf{Intents and Existing Rules.}  
We generate a dataset of 100 natural language intents for each network, with 50 manually crafted ones based on our NetOps interviews and another 50 generated by GPT-4o. 
``Permit'' and ``deny'' intents each consist of 50\% of the dataset.
Each intent uses endpoint names or prefixes from the corresponding SNMT and common applications extracted from real ACLs of a university network and a production cloud network. 
We include time ranges (e.g., weekdays) for half of the intents. 
Each attribute except action can be ``any'', but cases where source, destination, and application are all ``any'' are avoided. 
About 5\% intents have spelling errors in endpoint names in order to generate more realistic human inputs. 
Additionally, we handcraft 100 protect intents per network. \Cref{app:intent_example} has more details and examples about intents.
We also open-source intents, network topologies and SNMT in~\cite{xumi}.

We generate existing rules at each interface's ACL in the same way, using the endpoint prefixes information from SNMT of each network and applications from real ACLs. 
Each ACL also includes a default rule: ``deny all'' for \campus and ``permit all'' for the other two\footnote{Based on our review of empirical ACLs in university and production, campus networks prioritize security, denying all traffic and opening specific connections as needed; while cloud networks prioritize connectivity.}. 

\subsection{End-to-End Performance} \label{eval:end_to_end} 

\begin{table}[t]
     \centering
     \Huge
     \renewcommand{\arraystretch}{1.75}
     \resizebox{1.0\columnwidth}{!}{
\begin{tabular}{|c|ccc|cc|cc|c|}
\hline
 &
  \multicolumn{3}{c|}{\comprehend} &
  \multicolumn{2}{c|}{\conflict} &
  \multicolumn{2}{c|}{\textit{Deployment}} &
  \multirow{2}{*}{\begin{tabular}[c]{@{}c@{}}Total \\ Time (s)\end{tabular}} \\ \cline{2-8}
 &
  \begin{tabular}[c]{@{}c@{}}Avg. Acc.\\ w/o FB\end{tabular} &
  \begin{tabular}[c]{@{}c@{}}Avg. \# FB\\ Rounds\end{tabular} &
  Time (s) &
  \begin{tabular}[c]{@{}c@{}}\# Detected \\ Conflict ACLs\end{tabular} &
  Time (s) &
  \begin{tabular}[c]{@{}c@{}}\# Rule \\ Additions\end{tabular} &
  \begin{tabular}[c]{@{}c@{}}Decision\\ Time (s)\end{tabular} &
   \\ \hline
\sys &
  90\% &
  1.2 rounds &
  237.0 s &
  183 &
  55.0 s &
  285.6 &
  5.80 s &
  297.8 s \\ \hline
Baseline &
  18\% &
  17.2 rounds &
  3069.2 s &
  - &
  - &
  458.6 &
  0.1132 s &
  3069.31 s \\ \hline
\end{tabular}
}
    \vspace{5pt}
     \caption{
     End-to-end performance of \sys with \cloud and GPT-4o compared to a baseline inspired by prior works~\cite{lumi,netconfeval} and common practice. 
     }
     \label{tab:end_to_end}
     \vspace{-10pt}
\end{table}

We evaluate \sys's end-to-end performance in \cloud with GPT-4o and compare it to a baseline representing state-of-the-art ACL configuration practices:
\begin{itemize}
    \item Intent comprehension uses method similar to NetConfEval~\cite{netconfeval}, which only contains basic task description and network information (IR and SNMT) without any hallucination mitigators. We also invite NetOps to provide feedback on errors and re-prompt LLM until all intents are fulfilled, as in Lumi~\cite{lumi} and NetConfEval~\cite{netconfeval}. 
    \item Deployment follows common practices acknowledged by production NetOps (recall~\cref{moti:deployment}): ``permit'' rules are applied to all interfaces along all routing paths, and ``deny'' rules follow \textit{endpoint} deployment.
\end{itemize} 
Baseline excludes conflict detection, as existing methods like verifications are impractical for \textit{full} detection and may cause long or even infinite resolution loops (recall~\cref{moti:deployment}). 
In Baseline we assume all conflicts and resolved intents are given for deployment for fair comparison with \sys. 
Each interface randomly contains 20, 200, or 2000 rules, with 10 new ``permit'' and 10 ``deny'' intents per experiment. Evaluation results are averaged over 10 runs. 

Table~\ref{tab:end_to_end} summarizes the results. Overall, \sys completes configuration of 20 intents in 5 minutes on average, while Baseline takes over 10x longer. For Baseline comprehension, a simple user study with NetOps shows an estimated average time of $\sim$237s for 20 intents (with GPT-4o). We notice that most of the time in the Baseline workflow is spent on intent comprehension due to low accuracy and repeated human effort (Table~\ref{tab:end_to_end}). 
\sys enjoys much higher accuracy with hallucination mitigator design even without feedback, thus greatly reducing manual time. 
In Table~\ref{tab:end_to_end}, we omit manual resolution time for Baseline for fair comparison with \sys, as it does not involve such a process. 
In the deployment process, \sys reduces total number of deployed rules by $\sim$40\%, again highlighting the value of its optimization considering bottleneck and complementary-rule deployment.

\subsection{Intent Comprehension} \label{eval:comprehend}

We evaluate \comprehend using all original and protect intents, testing all four LLMs without feedback and GPT-4o with feedback until all intents are correctly comprehended. We also conduct ablation studies on each hallucination mitigator in \textit{system prompt} by measuring GPT-4o's performance when specific mitigators are excluded from \sys. 

\begin{figure}[t]
	\centering
	\includegraphics[width=1.0\columnwidth]{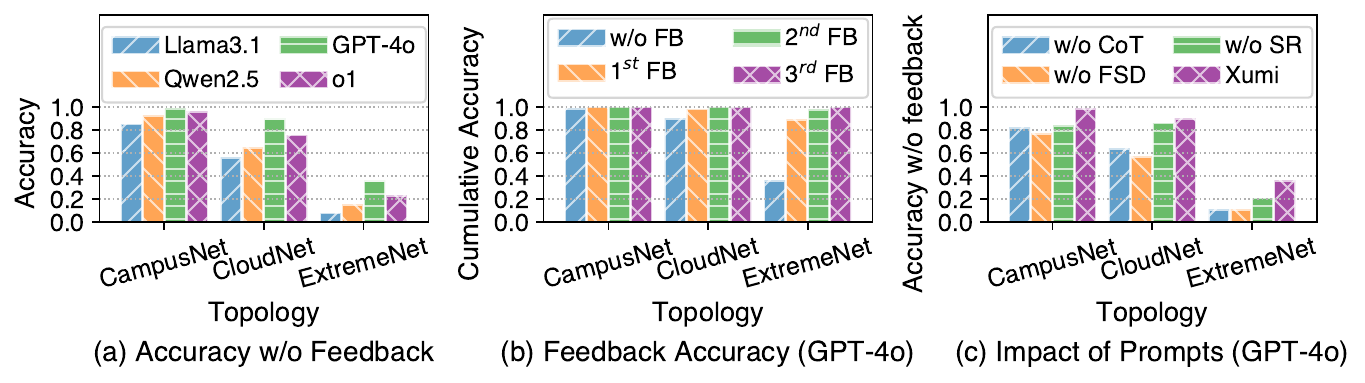}
    \vspace{-15pt}
	\caption{
    Comprehension accuracy across prompts and LLMs. (a) shows accuracy without NetOps' feedback across LLMs, (b) shows cumulative accuracy with feedback rounds using GPT-4o, and (c) evaluates specific hallucination mitigator under GPT-4o by excluding it from \textit{system prompt} and measuring accuracy without feedback. CoT refers to \textit{Chain-of-Thought Reasoning}, FSD to \textit{Few-Shot Demonstration}, and SR to \textit{Self-Reflection} in~\Cref{design:comprehend:hallucination}.
    }
 \label{fig:intent_metrics}
\end{figure}

\noindent\textbf{Comprehension Accuracy.}  
We first evaluate cases without NetOps feedback.  
As shown in Figure~\ref{fig:intent_metrics}(a), GPT-4o outperforms other LLMs with 98.5\% accuracy on \campus and 90\% on \cloud. 
Notably, o1 underperforms compared to GPT-4o, particularly on larger \cloud and \extreme, as it often produces partial outputs (e.g., mapping an entity to incomplete prefixes and interfaces). 
The smaller locally deployed models, Qwen and Llama, consistently underperform compared to GPT-4o and o1 due to their model size constraints, with Qwen outperforming Llama across all network scales. 
On \extreme, containing 2000+ entities and requiring $\sim$180k tokens (exceeding most LLMs' 128k context limit) which enforce \sys to apply iterative prompting (\Cref{sec:implement}), all models struggle with errors like incorrect interface identification or partial outputs mentioned above, leading to accuracy only with up to 36\% using GPT-4o. It suggests that the excessive length of SNMT overwhelms the models. We summarize common comprehension errors in~\Cref{app:error}. 

Figure~\ref{fig:intent_metrics}(b) shows that \sys achieves 100\% accuracy within 3 feedback rounds for all intents, including those in \extreme. 
On \extreme, single feedback from NetOps improves accuracy from 36\% to 89\%, while all error IRs on \campus and \cloud can be corrected within 1 and 2 rounds, respectively. This underscores that \sys's designs, including \textit{system prompt} and \textit{feedback prompt} (\cref{design:comprehend}), effectively improve accuracy with each LLM inference, ensuring comprehension is completed with minimal manual efforts, even in challenging scenarios beyond current LLM capabilities. 

We conduct ablation studies to evaluate the effectiveness of different hallucination mitigators (\Cref{design:comprehend:hallucination}). 
As shown in Figure~\ref{fig:intent_metrics}(c), compared to methods without specific mitigator, \sys's \textit{Few-Shot Demonstration} (FSD) achieves the highest accuracy improvement (26.5\%), followed by \textit{Chain-of-Thought Reasoning} (CoT) (22.5\%) and \textit{Self-Reflection} (SR) (11.2\%) averaged across all networks. 
Notably, all mitigators become more effective as network scale increases. 
For example, \sys's FSD improves accuracy by 1.27x over the method without it on \campus and by 3.27x on \extreme.

\begin{table}[t]
     \centering
     \Huge
     \renewcommand{\arraystretch}{1.35}
     \resizebox{1.0\columnwidth}{!}{

\begin{tabular}{|c|cccc|cccc|}
\hline
\multirow{2}{*}{Time (s)} & \multicolumn{4}{c|}{Initial Inference Time w/o Feedback} & \multicolumn{4}{c|}{Cumulative Feedback Time (GPT-4o)} \\ \cline{2-9} 
                          & Llama3.1-70B    & Qwen2.5-72B   & GPT-4o   & OpenAI o1   & w/o FB & 1$^{st}$ FB & 2$^{nd}$ FB & 3$^{rd}$ FB \\ \hline
\campus & 3.8 & 3.2 & 2.5 & 4.7 & 2.5 & 82.3 & - & - \\
\cloud & 8.7 & 6.9 & 5.2 & 11.4 & 5.2 & 71.7 & 96.0 & - \\
\extreme & 9.1 & 14.4 & 11.3 & 16.9 & 11.3 & 152.8 & 267.8 & 370.8 \\ \hline
\end{tabular}

}
    \vspace{5pt}
     \caption{Comprehension times of (1) Different LLMs' inference without feedback and (2) Cumulative feedback time across feedback rounds with human review, feedback, and GPT-4o inference time.}
     \vspace{-10pt}
     \label{tab:comprehend_time}
\end{table}

\noindent\textbf{Comprehension Time.} As shown in Figure~\ref{tab:comprehend_time}, all LLMs finish IR inference within 20 seconds without feedback, with GPT-4o being the fastest (6.3s on average), followed by Qwen, Llama, and o1 being the slowest (11.0s on average). 
Their inference times increase with topology size: a 112\%–-142\% increase from \campus to \cloud and 4\%–-116\% from \cloud to \extreme across four LLMs. 
Notably, Llama shows only a 4\% time increase from \cloud to \extreme but the lowest accuracy in \extreme since it always returns error IRs after processing the first SNMT slice without checking further slices during iterative prompting (\Cref{sec:implement}).
With all feedback, \sys with GPT-4o takes $\sim$3 minutes per intent on average and up to $\sim$8 minutes across networks. 
To conclude, GPT-4o achieves the best balance of speed and accuracy.

\subsection{Conflict Detection} \label{eval:detect}

We compare \sys with three methods that do not fully account for FPs: (1) \textit{Blind Overlapping} (\textit{BO}), a baseline method that naively identifies overlapping traffic between intents and independent existing rules with opposite actions as conflicts, ignoring both two FP cases; (2) \textit{BO+TMF}, which accounts for FPs caused by preceding rules based on \textit{BO}; and (3) \textit{BO+Path}, which accounts for FPs caused by infeasible paths based on \textit{BO}. 
Detection accuracy of an evaluated method is defined as the proportion of existing rules where the correct conflict flow set is identified for the new intent rule, across all existing ones installed at routing paths of intent's traffic.

We vary the number of existing rules per ACL from 20, 200, to 2000, all randomly generated. 
In each setting, 10 new intent rules are randomly selected from the comprehension results of our 100-intent dataset, and we report the results on average over 5 runs.

\begin{figure}[t]
	\centering
	\includegraphics[width=1.0\columnwidth]{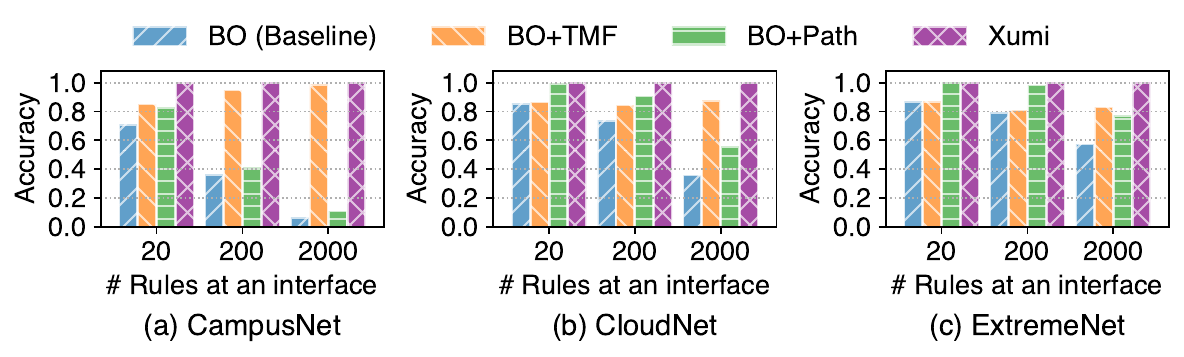}
    \vspace{-20pt}
	\caption{Conflict detection accuracy of \sys and methods ignoring FPs.}
 \label{fig:detection_metrics}
 \vspace{-15pt}
\end{figure}

\noindent\textbf{Detection Accuracy.} As shown in Figrue~\ref{fig:detection_metrics}, \sys achieves overall superior detection accuracy compared to other methods, with 3.33x, 1.15x, and 2.23x higher than \textit{BO}, \textit{BO+TMF}, and \textit{BO+Path}, respectively, averaged across all networks and rule number settings. 
\sys's accuracy gain increases with rule numbers, outperforming \textit{BO} and \textit{BO+Path} that do not consider TMF. 
For example, on average across three networks, \sys achieves 5.14x and 3.09x accuracy gains over \textit{BO}, and 3.57x and 2.24x over \textit{BO+Path} when each interface has 2000 rules compared to settings of 20 and 200. 
This is because more rules lead to higher overlap among existing rules in ACL, making TMF essential for accurate detection. Meanwhile, \textit{BO+TMF} consistently achieves accuracies near or above 80\% in all cases as shown in Figure~\ref{fig:detection_metrics}, reducing FPs and outperforming \textit{BO+Path} as rule number increases.

However, \sys's accuracy gain diminishes in larger networks with the same rule number. 
For example, with 2000 rules, \sys has 16.2x accuracy gain over \textit{BO} in \campus (Figure~\ref{fig:detection_metrics}(a)) but 1.75x in \extreme (Figure~\ref{fig:detection_metrics}(c)). 
This is because larger networks have more entities and flows, resulting in lower existing rule overlap for the same rule number setting. 
\textit{BO+Path} performs better than \textit{BO+TMF} in these cases, as seen with 20 or 200 rules in \cloud (Figure~\ref{fig:detection_metrics}(b)) and \extreme (Figure~\ref{fig:detection_metrics}(c)). 
In conclusion, neither TMF nor path validation alone achieves consistently high accuracy across all scenarios. By jointly addressing both, \sys overcomes their limitations and delivers robust and accurate detection.

\noindent\textbf{Detection Time.} As shown in Table~\ref{tab:detect_time}, \sys's detection time across its three phases increases with network scale and per-interface rule numbers. 
For the largest topology \extreme with 2000 rules per interface, detection for one new intent rule on all existing ones in an interface's ACL can be completed in 6 minutes. 
In the worst case, where an intent's traffic traverses all interfaces of \extreme, detection can still finish within 2 hours on our setup with three servers (192 concurrent threads each). 
\textit{TMF Calculation} and \textit{Intent Overlapping} (\cref{design:conflict:accurate}) dominate detection time, which is worthwhile due to their critical role in ensuring accuracy under high existing rule overlap, while \textit{Interface-Path Validation} is highly efficient by taking negligible time yet reduces FPs as shown in Table~\ref{tab:detect_time}.
Leveraging more concurrent threads across interfaces can further reduce overall detection time.

\begin{table}[t]
     \centering
     \huge
     \renewcommand{\arraystretch}{1.5}
     \resizebox{1.0\columnwidth}{!}{

\begin{tabular}{|c|ccc|ccc|ccc|}
\hline
\multirow{2}{*}{Time (s)} & \multicolumn{3}{c|}{\campus} & \multicolumn{3}{c|}{\cloud} & \multicolumn{3}{c|}{\extreme} \\ \cline{2-10} 
                   & 20     & 200    & 2000   & 20     & 200    & 2000   & 20    & 200   & 2000  \\ \hline
TMF Calculation    & 0.0021 & 0.0175 & 0.1819 & 0.1130 & 0.9076 & 13.44  & 2.581 & 17.90 & 248.0 \\
Intent Overlapping & 0.0007 & 0.0053 & 0.0398 & 0.0469 & 0.3119 & 3.926  & 1.189 & 6.420 & 72.22 \\
Interface-Path Validation & 0.0002   & 0.0002  & 0.0003  & 0.0006  & 0.0007  & 0.0010  & 0.0012   & 0.0014   & 0.0013  \\ \hline
Total Time         & 0.0030 & 0.0230 & 0.2220 & 0.1604 & 1.2202 & 17.367 & 3.771 & 24.32 & 320.2 \\ \hline
\end{tabular}

}
    \vspace{5pt}
     \caption{
     \sys's conflict detection time across three phases, averaged on the time of each intent rule detected at each interface's ACL. Detection across different interfaces and intent rules can be run in parallel.
     }
     \label{tab:detect_time}
     \vspace{-10pt}
\end{table}

\subsection{Deployment Optimization} \label{eval:deploy}

We evaluate \sys deployment optimization design. 
For ``permit'' intents, our baseline here is a \textit{Catch-All} method which simply deploys an intent on all potential interfaces with non-empty conflict flow sets. 
For ``deny'' intents, the baselines include: (1) \textit{Endpoint} method which deploys an intent on all source or destination gateway interfaces whichever has fewer ones; and (2) the \textit{Bottleneck} method which applies optimization only to identify bottleneck interfaces.

We conduct experiments by varying the average \textit{conflict ratio} across intents, which is defined as the fraction of interfaces (among those traversed by an intent's traffic) whose existing ACL conflicts with the intent. 
Thus higher conflict ratios indicate more interfaces are potential intent deployment targets (recall \Cref{design:deploy:eq}).  
In each setting, we first select 10 ``permit'' and 10 ``deny'' intents from intent dataset, then generate each ACL's existing rules randomly so that the average conflict ratio reaches the specified value. For each conflict here, its protect intent is randomly decided to be present or not.
Results are averaged over five runs for each setting.

\begin{figure}[t]
	\centering
	\includegraphics[width=1.0\columnwidth]{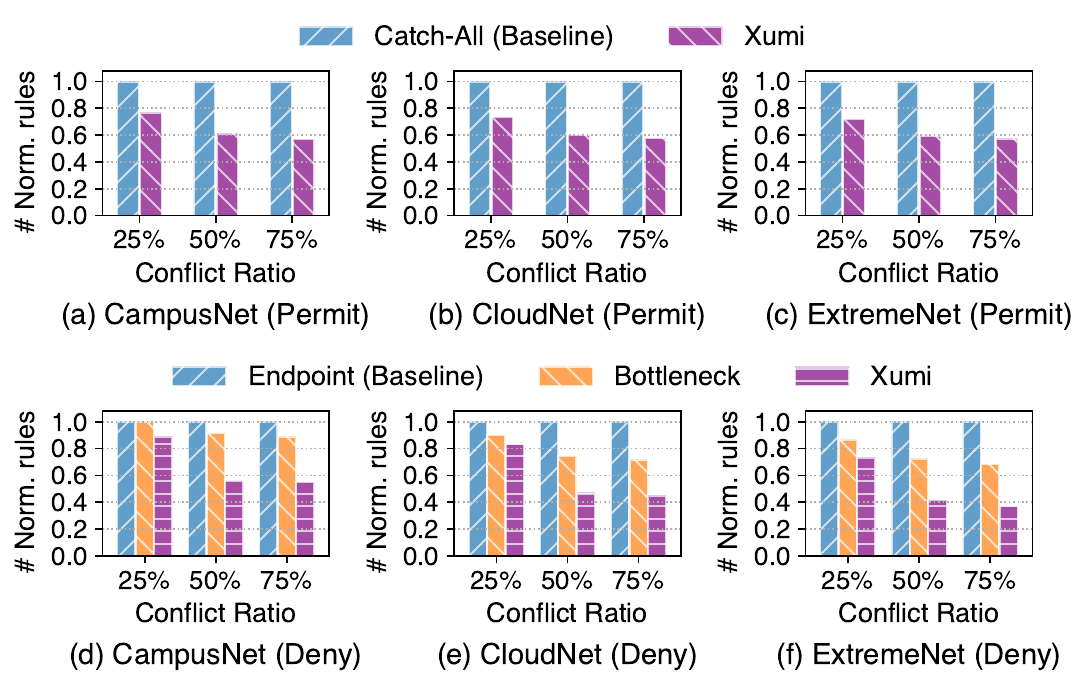}
    \vspace{-15pt}
	\caption{Normalized rule additions for different deployment optimization methods, with values normalized to corresponding baselines.}
 \label{fig:deploy_metrics}
 \vspace{-5pt}
\end{figure}

\noindent\textbf{Overall performance.} 
As shown in Figure~\ref{fig:deploy_metrics}, \sys reduces rule additions by 38.8\% on average across all settings. 
Specifically, \sys reduces ``permit'' rule additions by 36.3\% compared to \textit{Catch-All} (Figures~\ref{fig:deploy_metrics}(a)(b)(c)) and ``deny'' additions by 24.0\% over \textit{Bottleneck}, while \textit{Bottleneck} reduces them by 17.4\% over \textit{Endpoint} (Figures~\ref{fig:deploy_metrics}(d)(e)(f)).
Table~\ref{tab:deploy_time} shows \sys's decision time with varying network scale and conflict ratios. 
Even for \extreme with a 75\% conflict ratio, it can solve the optimization within 3 minutes, significantly better than the one-week manual configuration time reported in~\cite{jinjing}.

\noindent\textbf{Bottleneck Deployment.}  
We delve deeper into the impact of {bottleneck} deployment for the ``deny'' intents.  
As shown in Figure~\ref{fig:deploy_metrics}(d)(e)(f), as conflict ratio increases, {bottleneck} is more beneficial. 
This is because more conflicts lead to more routing paths that need to install new rules, increasing the likelihood of multiple intent paths traversing the same interface. 
Similarly, larger topologies also amplify the {bottleneck}'s effect. Compared to baseline, bottleneck saves 6.5\% rule additions in \campus (Figure~\ref{fig:deploy_metrics}(d)), but 24.2\% in \extreme (Figure~\ref{fig:deploy_metrics}(f)), respectively, as more complex routing in larger networks increases the chances of path overlap at interfaces.

\noindent\textbf{Complementary-Rule Deployment.}  
We analyze deeper into complementary rules' impact. 
Figure~\ref{fig:deploy_metrics} shows that using it greatly reduces rule additions compared to bottleneck deployment. 
As conflict ratios increase, it becomes more effective in reducing rule additions across all cases, because 
more conflicts mean more interfaces become potential deployment targets, and more intents may find their equivalences. 

As shown in Table~\ref{tab:deploy_time}, \sys solves optimizations quickly within 140s in the worst case. 
The solving time for ``permit'' is 23.4x higher than ``deny'' in the smaller \campus but only 0.263x in the larger \extreme, averaged across all conflict ratios. This is because ``permit'' requires multiple sequential per-interface optimizations, making it slower in smaller topologies, while ``deny'' involves a single optimization. In larger networks, however, global optimization for ``deny'' is more complex, taking more time than ``permit''.

\section{Discussion}
We discuss \sys's limitations and future works.

\noindent\textbf{Automated Routing Configurations.}  
\sys currently only focuses on automating ACL configurations to manage and filter network flows, assuming that routing configurations are correct and predefined. 
In contrast, routing protocols like BGP~\cite{bgp-cisco}, widely used in production networks, are far more complex due to their intricate protocol attributes and interdependencies. 
This complexity makes manual configuration error-prone, leading to potential misconfigurations that can disrupt desired reachability between endpoints and interfere with configurations of other flows. 
Inspired by \sys, a similar workflow could be applied to automate routing configurations, i.e., by understanding routing intents, detecting and resolving conflicts, and deploying correct configurations. 
And the unique complexity of routing protocols may present both novel challenges and exciting research opportunities for future work.

\noindent\textbf{Open-Ended Intents.} 
\sys currently supports expressing natural language intents tied to specific policies (i.e., ACLs) for configuration. However, in practice, NetOps may also want to propose open-ended intents that are not directly tied to a specific policy like ACL. Instead, they expect the system to infer appropriate configurations based on the current network state to fulfill the intent. 
For example, NetOps might request: ``Improve the WiFi signal in the technical department offices''. Addressing this intent could require selecting optimal WiFi frequencies and channels, adjusting transmission power, modifying bandwidth, or enabling QoS. 
The desired system should automatically diagnose the root causes of weak WiFi signals and identify the most suitable configuration adjustments for intent satisfaction. 
Handling such open-ended intents will be an exciting future work, presenting interesting research problems and new challenges.

\section{Related Work} \label{related}

We present related work not discussed before in \Cref{sec:intro} and \Cref{sec:moti}.

\noindent\textbf{Named-Entity Recognition (NER).} 
\sys's \comprehend resembles the well-established NER works~\cite{lumi,ner1,ner2,ner3}, including Lumi~\cite{lumi} from our network community. 
NER models are designed to extract ``names'', which are fragments of human expressions corresponding to specific entities, such as identifying the specifics of five key ACL attributes. 
However, applying NER approaches to our work, which often use the traditional models like BERT~\cite{bert}, may present challenges: These models typically stop at identifying the NL fragments, such as the endpoint and application names, without mapping them to the ACL specifics (e.g., prefixes and protocol ports) for practical use. 
Addressing diverse NL expressions and aligning them with ACL specifics requires tedious additional training or fine-tuning~\cite{vague1,vague2,correct1,correct2}.

\begin{table}[t]
     \centering
     \huge
     \renewcommand{\arraystretch}{1.6}
     \resizebox{1.0\columnwidth}{!}{
\begin{tabular}{|ccc|ccc|ccc|ccc|}
\hline
\multicolumn{3}{|c|}{\multirow{2}{*}{Time (s)}} &
  \multicolumn{3}{c|}{\campus} &
  \multicolumn{3}{c|}{\cloud} &
  \multicolumn{3}{c|}{\extreme} \\ \cline{4-12} 
\multicolumn{3}{|c|}{}                                                                   & 25\%   & 50\%   & 75\%   & 25\%   & 50\%   & 75\%   & 25\%   & 50\%   & 75\%   \\ \hline
\multicolumn{1}{|c|}{\multirow{2}{*}{\begin{tabular}[c]{@{}c@{}}``Permit''\\ Intents\end{tabular}}} &
  \multicolumn{1}{c|}{\multirow{2}{*}{\sys}} &
  EIS &
  0.0119 &
  0.0219 &
  0.0475 &
  0.9685 &
  2.268 &
  3.550 &
  36.96 &
  93.30 &
  152.2 \\
\multicolumn{1}{|c|}{} & \multicolumn{1}{c|}{}                                     & OPT & 0.0233 & 0.0148 & 0.0216 & 0.0288 & 0.0242 & 0.0338 & 0.0392 & 0.0417 & 0.0689 \\ \hline
\multicolumn{1}{|c|}{\multirow{3}{*}{\begin{tabular}[c]{@{}c@{}}``Deny''\\ Intents\end{tabular}}} &
  \multicolumn{2}{c|}{Bottleneck} &
  0.0016 &
  0.0027 &
  0.0036 &
  0.0725 &
  0.0847 &
  0.0946 &
  0.4951 &
  0.5073 &
  0.5538 \\ \cline{2-12} 
\multicolumn{1}{|c|}{} & \multicolumn{1}{c|}{\multirow{2}{*}{\sys}} & EIS & 0.0162 & 0.0215 & 0.0306 & 1.760  & 3.174  & 4.080  & 49.29  & 104.7  & 138.93 \\
\multicolumn{1}{|c|}{} & \multicolumn{1}{c|}{}                                     & OPT & 0.0017 & 0.0031 & 0.0044 & 0.0727 & 0.0812 & 0.0939 & 0.5125 & 0.5638 & 0.6144 \\ \hline
\end{tabular}
}
    \vspace{5pt}
     \caption{Deployment plan decision time with \sys's different optimization observations and conflict ratios (25\%, 50\%, 75\%). \sys's time includes ``EIS'' for \textit{Equivalent Intent Set} computation and ``OPT'' for optimization. 
     }
     \vspace{-15pt}
     \label{tab:deploy_time}
\end{table}

\noindent\textbf{Network Synthesizer.} 
Extensive research on network synthesizers~\cite{netcomplete,Aura,cegs,nassim-sigcomm22,cosynth} has focused on filling configuration templates with parameters according to specific intents, but these intents are often described at a low level, requiring NetOps to handcraft detailed attributes such as prefixes and routers, thus limiting automation. 
For example, Aura~\cite{Aura} requires NetOps to specify paths manually for a routing configuration and automates only simple tasks like determining global BGP communities. 
Recent works like CEGS~\cite{cegs} and COSYNTH~\cite{cosynth} leverage LLMs to interpret high-level intents and select appropriate templates for certain devices, while Nassim~\cite{nassim-sigcomm22} uses language models to translate parameters between vendor templates. 
However, these methods still suffer from the limitations of traditional synthesizers in attribute handcrafting. Additionally, synthesizers focus solely on configuration correctness, ignoring conflict resolution and deployment simplicity, which are issues that our work addresses with clear advantages.

Although these synthesizers are not suitable for \sys's ACL configuration pipeline, we can still leverage their template-based approach to convert \sys's ACL rule quintuple into deployable scripts for realistic devices, since \sys is not designed to directly generate ACL scripts from natural language intents. We assumes templates are available as demonstrated in prior works~\cite{robotron2016,openconfig-ref,apstra-gdb,nassim-sigcomm22}.

\section{Conclusion} \label{sec:conclusion} 
We propose \sys, a novel automated ACL configuration system that helps translate NetOps' natural language intents into router configurations with minimal manual intervention. Using advanced prompt designs to tackle network-specific information and LLM hallucinations, \sys achieves high comprehension accuracy in deriving valid ACL rules from intents with LLMs, significantly reducing manual feedback. Its proactive conflict detection accurately identifies all conflicts for resolution, avoiding risky deploy-then-fix process, while deployment optimizations reduce rule additions and simplify maintenance. Evaluation results demonstrate that \sys effectively automates entire ACL configuration pipeline, and all its module designs perform efficiently as expected.

\clearpage
\bibliographystyle{plain}
\bibliography{reference}

\clearpage
\appendix
\section*{Appendices}
\addcontentsline{toc}{section}{Appendix}

\section{Comprehension Prompts and Human Interaction Example} \label{app:prompt_example}

\begin{figure}[ht]
	\centering
\includegraphics[width=1.0\columnwidth]{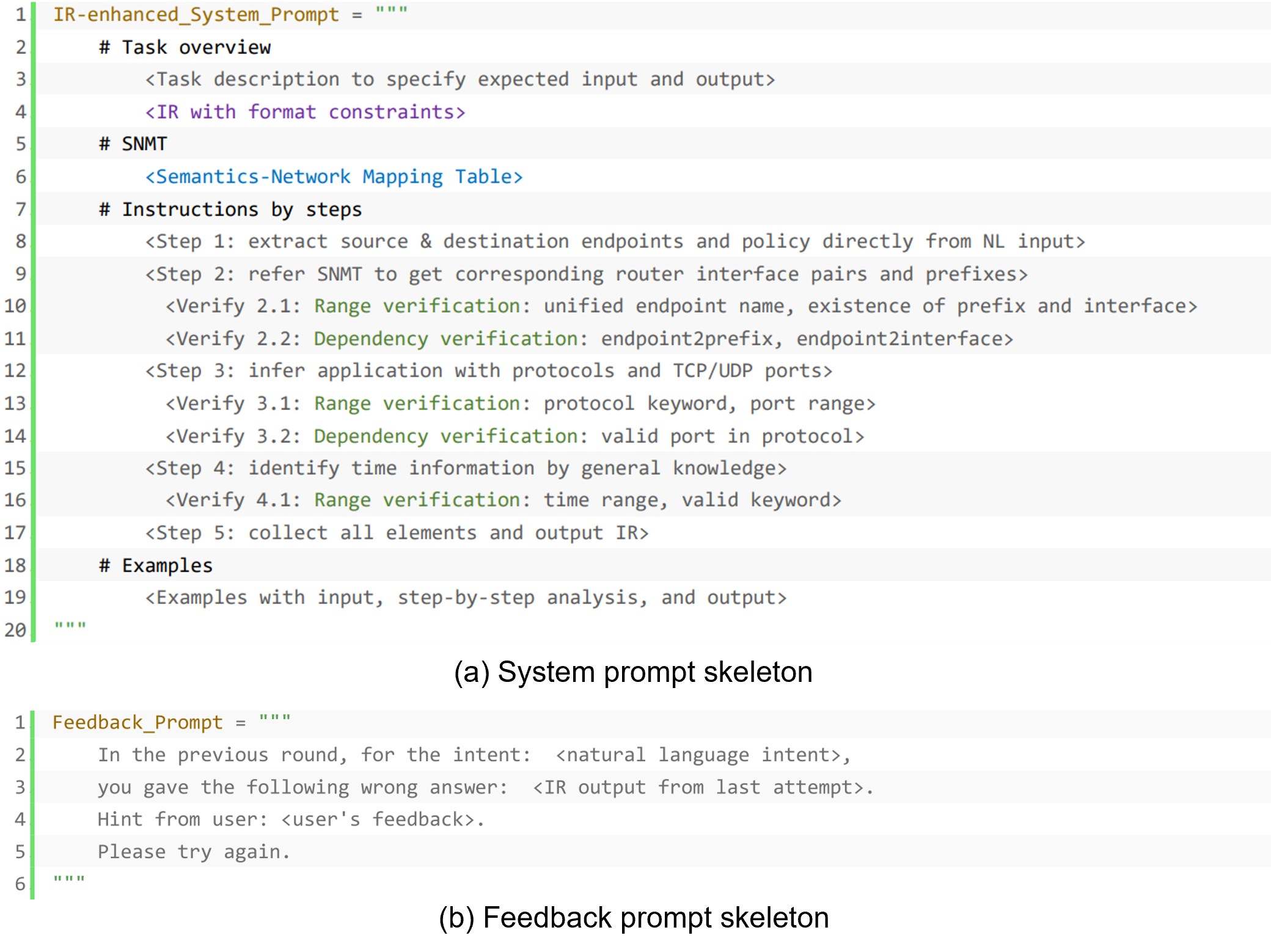}
	\caption{The skeleton of the prompts used in \comprehend. Note that ``Step 1–5'' corresponds to the \textit{Chain-of-Thought Reasoning} prompts, ``Examples'' represent the \textit{Few-Shot Demonstration} prompts, and ``Verify x.x'' refers to the \textit{Self-Reflection} prompts.}
 \label{fig:prompt_skeleton}
\end{figure}

\noindent\textbf{Prompt Skeletons.} As discussed in \cref{design:comprehend:hallucination}, we use a system prompt that guides LLMs to complete the \textit{Intent Comprehension} task and a feedback prompt that requires NetOps to check IR and provide human feedback for any comprehension errors. Figure~\ref{fig:prompt_skeleton} shows the skeleton of two prompts.

More specifically, the system prompt is divided into several parts as shown in Figure~\ref{fig:prompt_skeleton}. First, in the task overview, we define the organization of IR and the format constraints for each item. Second, SNMT is then loaded for the LLM's reference.  
Third, the step-by-step instructions are provided for guidance, where \textit{Chain-of-Thought Reasoning} and \textit{Self-Reflection} strategies are applied together. 
At each step of instruction, the LLM is asked to focus on extracting the specifics of one ACL attribute and verify the element range and dependencies of each extracted element results. 
It moves on to the next step only if all verification passes, and finally organizes and formats the entire IR at Step 5. Lastly, \textit{Few-Shot Demonstration} is provided, each example includes the sample intent and its ground truth IR, with a further step-by-step explanation about why the ground truth is correct. 
In practice, edge cases are found particularly useful for better overall performance. 
All specific prompts used in our evaluation are open-sourced at~\cite{xumi}.

The feedback prompt skeleton is shown in Figure~\ref{fig:prompt_skeleton}(b), and all its details are discussed in~\Cref{design:comprehend}.

\begin{figure}[t]
	\centering
	\includegraphics[width=1.0\columnwidth]{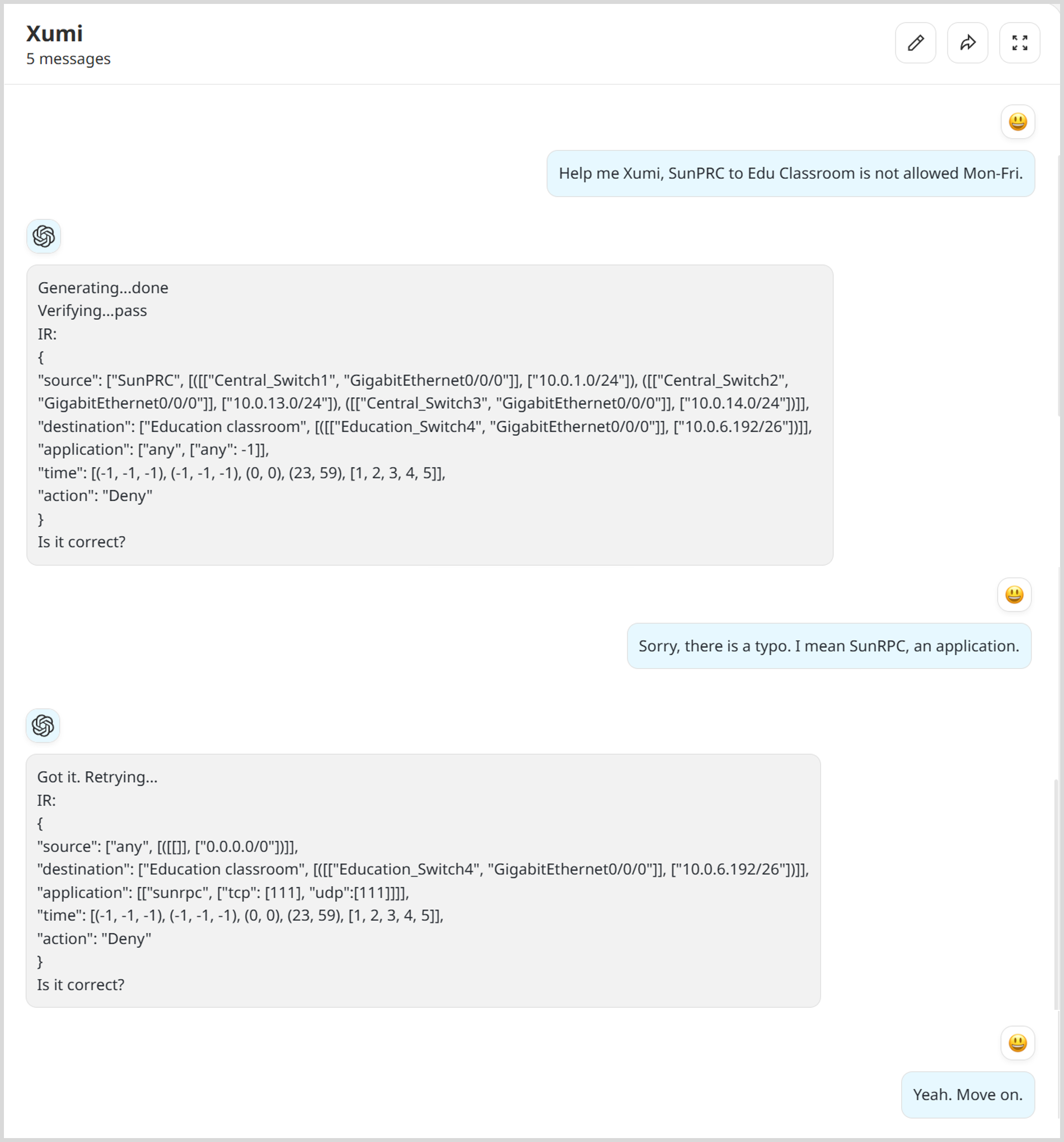}
	\caption{An interaction example under web-chat-based demo. NetOps are invited to review the IR output and provide feedback if any errors are detected.}
 \label{fig:interaction_example}
\end{figure}

\noindent\textbf{NetOps Interaction.} The NetOps can interact with \sys just like talking to a normal chat robot, as illustrated in Figure~\ref{fig:interaction_example}. 
The agent is always initialized with a system prompt before it starts. Then, NetOps can specify an intent, making the agent to be triggered to generate an IR for manual review. 
If any errors are identified, NetOps provides manual feedback, which the agent uses to regenerate result IR. 
This feedback loop continues until NetOps confirms correctness. 
Finally, an IR of an intent is transformed into corresponding ACL rules of the intent as discussed in~\Cref{design:comprehend:generation}, facilitating conflict detection and rule deployment.

\begin{table*}[]
\centering
     \Huge
     \renewcommand{\arraystretch}{1.5}
    \resizebox{1.0\linewidth}{!}{%
\begin{tabular}{c|c|c|c|l}
\hline
Network & \begin{tabular}[c]{@{}c@{}}Type\\ (50\% Permit/Deny)\end{tabular} & Number & \#chars & \multicolumn{1}{c}{Example} \\ \hline
\multirow{2}{*}{\textit{CampusNet}} & Original intent & 100 & 64 & Biz Admin Classroom should always be able to communicate with Edu Office. \\
 & Protect intent & 100 & 112 & Protect the HTTPS flow from 10.0.7.0/24 in Law to 10.0.6.0/24 in Education over TCP port 443, as defined in the current configuration. \\ \hline
\multirow{2}{*}{\textit{CloudNet}} & Original intent & 100 & 65 & Authorize SunPRC traffic from East Asia DC5 to any destination on weekends. \\
 & Protect intent & 100 & 107 & OSPF routing traffic from 10.123.0.0/16 to 11.27.0.0/16 must be protected on weekends. \\ \hline
\multirow{2}{*}{\textit{ExtremeNet}} & Original intent & 100 & 65 & All traffic from NZ North C to Japan West F is denied Mon-Fri. \\
 & Protect intent & 100 & 107 & UDP traffic from Sweden Central C DC2 to Central India D DC2 must be protected from any changes. \\ \hline
\end{tabular}
}
\vspace{0.1pt}
 \caption{Detailed information of \textit{Intent Comprehension} dataset with examples.}
 \label{tab:understanding_dataset_detailed_info}
\end{table*}

\section{Examples of Intents} \label{app:intent_example}
Table~\ref{tab:understanding_dataset_detailed_info} provides detailed information about the intent dataset used for \textit{Intent Comprehension} evaluation. 
For each network in~\Cref{eval}, 100 original intents and 100 protect intents are generated. 
For each network, entities are designed to be meaningful and may contain other entities. 
For example in Table~\ref{tab:understanding_dataset_detailed_info}, ``Edu Office'' in the first example (first line) is included in ``Education'' as a whole university department in the second example (second line). 
Endpoints (source/destination) in intent can contain prefixes rather than endpoint names only. For example, there are prefixes-only intents like ``Protect the HTTPS flow from 10.0.7.0/24 in Law to 10.0.6.0/24 in Education over TCP port 443'' and ``OSPF routing traffic from 10.123.0.0/16 to 11.27.0.0/16 must be protected on weekends''. 
To enhance linguistic diversity, the datasets incorporate abbreviations, acronyms, and typos. Approximately 5\% of the data contains intentional typos, such as ``SunPRC'' for ``SunRPC'', while another 5\% includes abbreviations, such as ``Biz Admin Classroom'' for ``Business Administration Classroom'' or ``NZ North C'' for ``New Zealand North C,'' in Table~\ref{tab:understanding_dataset_detailed_info}. 

These datasets are all open-sourced at~\cite{xumi} to facilitate further research within the community.

\section{Comprehension Error Causes Analysis} \label{app:error}
For one-shot comprehension across LLMs, the most common error is incorrect interface/prefix mapping (63\%), with varying behaviors among models. For instance, in \cloud, where a region endpoint should be mapped to 5 interface-prefix tuples (2 interfaces per prefix), o1 returns only the first tuple, while Llama includes all 5 tuples but only the first interface for each. 
Qwen, limited by its 32k context length, struggles with tackling SNMT inputs beyond the first 1000 lines, leading to incorrect mappings. 

Wrong endpoint name errors (34\%) are frequent when intents provide only prefixes for endpoints, with Qwen and Llama more prone to this mistake. 
Errors related to applications or time are rare (3\%) and include issues like failing to identify typos (e.g., incorrectly writing "sunprc" as "sunrpc") or unnecessarily generating absolute time references (e.g., ``year 2023 or 2024''). Llama occasionally introduces formatting errors, such as using square brackets [] instead of parentheses (). Notably, no errors related to wrong actions were observed across any model.

\section{Bitarray-Based Flow Set Operations} \label{app:bitarray}

\noindent\textbf{Basic Set Operations with Bitarray.} 
As described in~\Cref{sec:implement}, each flow set is represented as a bitarray, where each bit indicates the presence of a specific flow in the network. All flows, characterized by combinations of attributes such as \textit{source}, \textit{destination}, and \textit{application}, are recorded in the \textit{global flow information table}.  
We leverage the bitwise operations of \textit{Bitarray} to perform the typical flow set operations required by \sys. For instance, suppose $A$ and $B$ are two flow sets represented in \textit{Bitarray} format. The \textit{intersection} ($\cap$) of $A$ and $B$ is computed as $A \land B$, the \textit{difference} ($-$) as $A \land \neg B$, and the \textit{union} ($\cup$) as $A \lor B$. Additionally, $A$ is determined to be a \textit{subset} ($\subseteq$) of $B$ if their \textit{difference} is empty, i.e., $(A \land \neg B).\text{count}(1) = 0$.  

These Bitarray-based flow set operations cover all operations required by \sys. For example, \textit{intersection} is used in \textit{Intent Overlapping} (\Cref{design:conflict:accurate}), \textit{difference} is used in \textit{TMF Calculation} (\Cref{design:conflict:accurate}), and \textit{union} and \textit{subset} are used in equivalent condition calculations (\Cref{design:deploy:eq}).

\noindent\textbf{Efficient Indexed Flow Set Extraction of A Rule.} 
Since we now conduct flow set operations using indexed flow sets, each rule must be expanded into a set of indexes corresponding to its flows for these operations. At first glance, this requires enumerating all flows governed by the rule and finding their indexes in the \textit{global flow information table}. To achieve this, we enumerate all possible values (or specifics) for each attribute of the rule and generate combinations of these specifics across attributes to identify individual flows. For example, the rule ``deny 10.0.0.0/31 $\rightarrow$ 11.0.0.0/31 HTTP'' has its source attribute ``10.0.0.0/31'' expanded into ``10.0.0.0/32'' and ``10.0.0.1/32,'' while its destination attribute ``11.0.0.0/31'' expands into ``11.0.0.0/32'' and ``11.0.0.1/32,'' with the application attribute being HTTP. By combining these specifics, we generate four flows: ``10.0.0.0/32 $\rightarrow$ 11.0.0.0/32 HTTP'', ``10.0.0.1/32 $\rightarrow$ 11.0.0.0/32 HTTP'', ``10.0.0.0/32 $\rightarrow$ 11.0.0.1/32 HTTP'', and ``10.0.0.1/32 $\rightarrow$ 11.0.0.1/32 HTTP''. Next, we locate these flows in the \textit{global flow information table} to determine their indexes. For instance, if the global table assigns indexes 1, 2, 3, and 4 to these flows, their corresponding Bitarray representations would be ``1000'', ``0100'', ``0010'' and ``0001'', respectively, assuming only these four flows exist in the network. However, directly enumerating all flows for a rule and searching for their indexes in a large table is highly inefficient.

To improve efficiency, we use a precomputed index set for each specific value of each attribute, stored in the \textit{global flow information table}. For example, for the specific source prefix ``10.0.0.0/32'', the precomputed index set includes all indexes of flows related to this prefix. In this case, ``10.0.0.0/32'' corresponds to the flows ``10.0.0.0/32 $\rightarrow$ 11.0.0.0/32 HTTP'' (index 1) and ``10.0.0.0/32 $\rightarrow$ 11.0.0.1/32 HTTP'' (index 3), so its index set is \{1, 3\}, represented as the Bitarray ``1010''. Using these precomputed index sets, we can efficiently compute the flow set for a rule. First, we enumerate all specifics of an attribute and take the union of their precomputed index sets. For example, in the rule ``deny 10.0.0.0/31 $\rightarrow$ 11.0.0.0/31 HTTP'', the source prefix ``10.0.0.0/31'' expands into the specifics ``10.0.0.0/32'' (Bitarray ``1010'') and ``10.0.0.1/32'' (Bitarray ``0101''), and the union of these sets is ``1010 $\lor$ 0101 = 1111,'' representing all flows related to the source. Similarly, we compute the flow sets for the destination and application attributes. Finally, we compute the intersection of the resulting flow sets to obtain the final indexed flow set for the rule. For example, in the rule ``deny 10.0.0.0/31 $\rightarrow$ 11.0.0.0/31 HTTP,'' the source, destination, and application attributes all yield the Bitarray ``1111'' in this example, so the final flow set of the rule is $1111 \land 1111 \land 1111 = 1111$.

\end{document}